\documentclass[11pt]{article}

\usepackage{fancyhdr}

\usepackage{geometry}

\usepackage{bm}
\usepackage{bbm}
\usepackage{float}
\usepackage{tikz}
\usepackage{amsthm}
\usepackage{amssymb}
\usepackage{subfig}
\usepackage[round]{natbib}
\usepackage{enumerate}
\usepackage{longtable}
\usepackage{mathtools,cancel}
\usepackage{url}
\usepackage[ruled,vlined,linesnumbered,lined,boxed,commentsnumbered]{algorithm2e}
\usetikzlibrary{positioning}
\usetikzlibrary{graphs} 
\usetikzlibrary[graphs]
\usepackage{multirow}
\usepackage{lscape}
\usepackage[english]{babel}
\usepackage{amsmath}

\usepackage[colorlinks,citecolor=blue,urlcolor=blue,filecolor=blue,backref=page]{hyperref}

\usepackage{soul}

\renewcommand{\Re}{\mathbb{R}}
\newcommand\diag{\mathrm{diag}}

\newcommand{\vat}{\mathbb{E}}
\newcommand{\var}{\mathbb{V}\!\mathrm{ar}}

\newcommand{\bc}{\begin{center}}
	\newcommand{\ec}{\end{center}}

\newcommand{\bit}{\begin{itemize}}
	\newcommand{\eit}{\end{itemize}}

\newcommand{\be}{\begin{eqnarray*}}
	\newcommand{\ee}{\end{eqnarray*}}
\newcommand{\ben}{\begin{eqnarray}}
\newcommand{\een}{\end{eqnarray}}

\newcommand{\g}{\,\vert\,}

\newcommand{\D}{\mathcal{D}}

\newcommand{\M}{\mathcal{M}}

\newcommand{\N}{\mathcal{N}}
\newcommand{\W}{\mathcal{W}}

\newcommand{\pa}{\mathrm{pa}}
\newcommand{\fa}{\mathrm{fa}}

\newcommand{\bzero}{\bm{0}}

\newcommand{\bA}{\bm{A}}

\newcommand{\bD}{\bm{D}}

\newcommand{\bI}{\bm{I}}

\newcommand{\bL}{\bm{L}}
\newcommand{\bM}{\bm{M}}

\newcommand{\bT}{\bm{T}}
\newcommand{\bU}{\bm{U}}
\newcommand{\bX}{\bm{X}}

\newcommand{\bb}{\bm{b}}

\newcommand{\bu}{\bm{u}}

\newcommand{\bx}{\bm{x}}
\newcommand{\by}{\bm{y}}

\newcommand{\bSigma}{\bm{\Sigma}}

\newcommand{\bOmega}{\bm{\Omega}}
\newcommand{\bGamma}{\bm{\Gamma}}
\newcommand{\bTheta}{\bm{\Theta}}

\newcommand{\bgamma}{\bm{\gamma}}

\newcommand{\btheta}{\bm{\theta}}
\newcommand{\bmu}{\bm{\mu}}

\newcommand{\ba}{\bm{a}}

\usepackage{tikz}
\usetikzlibrary{positioning}

\RequirePackage[OT1]{fontenc}
\RequirePackage{amsthm,amsmath}

\newtheorem{prop}{Proposition}[section]

\theoremstyle{plain}

\newcommand{\black}{\color{black}}

\SetKwInput{KwInput}{Input}
\SetKwInput{KwOutput}{Output}

\usepackage[linesnumbered,ruled,vlined]{algorithm2e}

\usepackage{xr}
\begin{document}

\title{Bayesian causal inference in probit graphical models}
\author{Federico Castelletti \& Guido Consonni
\footnote{Department of Statistical Sciences, Universit\`{a} Cattolica del Sacro Cuore, Milan, federico.castelletti@unicatt.it, guido.consonni@unicatt.it}}

\date{}

\maketitle


	\begin{abstract} 
		We consider a binary response which is potentially affected by a
		set of continuous variables.
		Of special
		interest is the causal effect on the response due to an
		intervention on a specific variable.
		The latter can be meaningfully determined on the basis of observational data
		through suitable assumptions on the data generating mechanism.
		In particular we assume that the joint distribution obeys the conditional independencies (Markov properties) inherent in a Directed Acyclic Graph (DAG), and the DAG is given a causal interpretation through the notion of interventional distribution.
		We propose a DAG-probit model where 
		the response is generated by discretization through a random threshold of a continuous latent variable and the latter, jointly with the remaining continuous variables, has a distribution belonging to a zero-mean Gaussian model whose covariance matrix is constrained to satisfy the Markov properties of the DAG. Our model leads to a natural definition of causal effect conditionally on a given DAG.
		Since the DAG which generates the observations  is unknown, we present an efficient MCMC algorithm whose target is the posterior distribution on the space of DAGs, the Cholesky parameters of the concentration matrix, and the threshold linking the response to the latent.
		%
		%
		Our end result is a Bayesian Model Averaging estimate of the causal effect
		which incorporates parameter, as well as model, uncertainty.
		The methodology is assessed using simulation experiments and applied to a
		gene expression data set originating from breast cancer stem cells.
	\end{abstract}


\section{Introduction}



We consider  a system of random quantities, which  includes  a binary response as well  as a collection of continuous variables, and address the problem of determining the causal effect on the response due to an intervention on a given variable.
A causal question involves  the data generating mechanism after an intervention is applied to the system, and must be carefully distinguished from   traditional conditioning of probability theory \citep[Section 2.4]{Pearl:2009:Surveys}.
The gold standard for addressing causal questions is represented by  randomized controlled experiments; the latter however are not always available because they may be unethical, infeasible, time consuming or
expensive \citep{Maathuis:Nandy:Review}.
By contrast, \emph{observational} data,  that is observations produced without exogenous perturbations of the system, are widely available and often plentiful.
The challenge is then  to infer causal effects  based on observational data alone.
To achieve this goal,  it is crucial to set up a a suitable conceptual framework  which is able to address causal questions; in particular the notion of  \emph{joint distribution} for a collection of random variables can only address concepts linked to association,  so much so that, by converse,  \lq \lq a causal concept is any relationship that
cannot be defined from the distribution alone\rq \rq{} \citep[Section 2]{Pearl:2009:Surveys}.

A very useful framework to bridge the gap between the observational and the experimental domains is represented by
the  Directed Acyclic Graph (DAG), or its allied concept of Structural Equation Model (SEM); see \citet{Pearl:1995} and \citet{Pear:2000}.
DAGs  have been extensively used to construct statistical models embodying  conditional independence relations \citep{Laur:1996}.
Applications are numerous especially in genomics; see for instance  \citet{Frie:2004} and \citet{Frie:Koll:2003}.
With observational data,    conditional independence relations will drive inference  on DAG and parameter space.
On the other hand, the additional    syntax and semantics  of \emph{causal} DAGs \citep{Pear:2000} will be instrumental to define the notion of causal effect.

As in standard probit regression  \citep{Albert:Chib:1993},
we  assume that  the observable binary response is the result of a discretization of a continuous \emph{latent} variable.
Next, for a given DAG,  we model all continuous random variables, along with the latent,  as a multivariate Gaussian family satisfying the corresponding Markov property.
We call the resulting setup  a  \emph{DAG-probit} model, and
\black
provide a
definition of causal effect on the response which is predicated on a given DAG through the notion of \emph{interventional} distribution \citep{Pear:2000}.
However the structure of the DAG is usually unknown,  and this must be taken into account
because different DAGs will typically induce distinct causal effects; see the review paper \citet{Maathuis:Nandy:Review}
and
\cite{Caste:Cons:2020} for a Bayesian approach.


In this work we extend the notion of interventional distribution and causal effect \citep{Pear:2000,Maat:etal:2009} to DAG-probit models.
Specifically, we propose a Bayesian method which jointly performs DAG-model determination
as well as inference of causal effects in the presence of a binary response.
From a computational viewpoint we introduce an MCMC scheme
to sample from the joint posterior of models (DAGs) and model-dependent parameters (causal effects) which we implement through an efficient PAS algorithm \citep{Godsill:2012}.
\black
The rest of the paper is organized as follows.
In Section \ref{sec:model:formulation} we review DAG-Gaussian models and define the DAG-probit model.
In Section \ref{sec:causal:effects} we present the structure of the interventional distribution in its general form, then specialize it to the Gaussian case,
and finally extend the definition of causal effect to
DAG-probit models.
Section \ref{sec:bayesian:inference} presents our Bayesian methodology with particular emphasis on  priors for  model parameters. An MCMC algorithm for posterior inference
on models, parameters and hence causal effects
is presented in Section \ref{sec:MCMC}. We evaluate the proposed methodology through simulation studies in Section \ref{sec:simulations},  and then apply it to a data set on gene expression measurements derived from breast cancer stem cells
(Section \ref{sec:application}). Finally a few points for discussion are presented in Section \ref{sec:discussion}. Some theoretical results as well as additional simulation outputs are reported in the Supplementary material \citep{caste:cons:suppl:BA}.

\section{Model formulation}
\label{sec:model:formulation}

In this section
we first provide some background material
on Gaussian DAG-models with special  emphasis on  their parameterization (Section \ref{subsec:Gaussian:DAG:models}).
Next we present our DAG-probit model
(Section \ref{subsec:DAG:probit:models}).
Both sections deal with the likelihood, while
choices of prior distributions   
are discussed in Section  \ref{sec:bayesian:inference}.

\subsection{Gaussian DAG-models}
\label{subsec:Gaussian:DAG:models}

%
%

\black

Let $\D=(V,E)$ be a DAG, where $V=\{1,\dots,q\}$ is a set of vertices (or nodes) and $E\subseteq V \times V$ a set of edges whose elements are $(u,v)\equiv u \rightarrow v$, such that if $(u,v) \in E$ then $(v,u) \notin E$. In addition, $\D$ contains no cycles, that is paths of the form $u_0 \rightarrow u_1 \rightarrow \dots \rightarrow u_k$ where $u_0 \equiv u_k$. For a given node $v$, if $u\rightarrow v \in E$ we say that $u$ is a \emph{parent} of $v$ (conversely $v$ is a \emph{child} of $u$). The parent set of $v$ in $\D$ is denoted by $\pa_{}(v)$, the set of children by $\textnormal{ch}_{}(v)$. Moreover, we denote by $\fa(v)=v\cup \pa(v)$ the \textit{family} of $v$ in $\D$.
For further theory and notation on DAGs we refer to \cite{Laur:1996}.


We consider a collection of random variables $(X_1,\dots,X_q)$ and  assume that their joint probability density function $f(\bx)$ is Markov w.r.t. $\D$, so that it admits the following factorization
\ben
\label{eq:DAG:facorization}
f(x_1,\dots,x_q)
=
\prod_{j=1}^{q}f(x_j\g\bx_{\pa(j)}).
\een
In this section,  as well as in Section \ref{sec:causal:effects},  we reason \textit{conditionally} on a given DAG $\D$ without an explicit conditioning in the notation  we use.
In Section \ref{sec:bayesian:inference} we will instead deal with model (DAG) uncertainty and will reinstate $\D$ in our notation.

If the joint distribution is Gaussian with mean equal to zero, we write
\ben
X_1,\dots,X_q\g\bOmega_{} \sim \N_q(\bzero,\bOmega_{}^{-1}),
\, \bOmega \in \mathcal{P}_{\D}
\een
where $\bOmega_{}=\bSigma_{}^{-1}$ is the precision matrix, and
$\mathcal{P}_{\D}$ is the space of symmetric positive definite (s.p.d.) precision matrices Markov w.r.t. $\D$.
%
%
For a Gaussian DAG-model factorization \eqref{eq:DAG:facorization} becomes
\ben
\label{eq:Gaussian:DAG:facorization}
f(x_1,\dots,x_q\g \bOmega)
=
\prod_{j=1}^{q}
d\,\N(x_j \g \mu_{j}(\bx_{\pa(j)}), \sigma_j^2),
\een
where
$d\,\N(\cdot \g \mu,\sigma^2)$
denotes the normal density having mean $\mu$ and variance $\sigma^2$.
The assumption of normality guarantees that the model if \emph{faithful} to the DAG,
that is all and only those conditional independencies emboded in the joint distribution can be read-off from the graph using the Markov property;
see also \citet{Spir:Glym:Sche:2000}.

Without loss of generality, we assume a \emph{parent ordering} of the nodes which numerically labels the variables so that $i > j$ whenever $j$ is a child of $i$.
A parent ordering always exists, although it is not unique in general.
We also remark that a parent ordering is specific to any given DAG under consideration and
may change if alternative DAGs are entertained.
Importantly, it only represents a convenient device to specify a prior on the parameter space; see Section \ref{sec:bayesian:inference} where we also show that the choice of the parent ordering does not affect the prior assigned to model parameters.

Moreover, we declare node 1, which cannot have children, to be the (latent) outcome variable.
Equation \eqref{eq:Gaussian:DAG:facorization}
can  be also written as a \emph{structural equation model}
\ben
\label{eq:structural:equation:model}
\bL^\top(X_1,\dots,X_q)^\top = \boldsymbol{\varepsilon},
\een
where because of the assumed parent ordering $\bL$ is a $(q,q)$ \emph{lower-triangular} matrix of coefficients, $\bL = \{\bL_{ij}, i \ge j\}$, such that $\bL_{ij}\ne 0$ if and only if $i \rightarrow j$ and $\bL_{ii}=1$. Moreover, $\boldsymbol{\varepsilon}$ is a $(q,1)$ vector of error terms,
$\boldsymbol{\varepsilon} \sim \N_q(\boldsymbol{0},\bD)$, where $\bD=\diag(\boldsymbol{\sigma}^2)$ and
$\boldsymbol{\sigma}^2$ is the $(q,1)$ vector of \emph{conditional} variances whose $j$-th element is $\sigma^2_j = \var(X_j\g\bx_{\pa(j)},\bOmega)$.
From \eqref{eq:structural:equation:model} it follows   that
\ben
\label{eq:cholesky:fact}
\bOmega=\bL\bD^{-1}\bL^{\top}.
\een
We refer to Equation \eqref{eq:cholesky:fact} as the \emph{modified Cholesky decomposition} of $\bOmega$. Let now $\prec j \succ \, =\pa(j)$ and  $\prec j\,]=\pa(j)\times j$.
Representation  \eqref{eq:cholesky:fact} induces a re-parametrization of $\bOmega$ 
in terms of the  Cholesky parameters
$\left\{
(\sigma_j^2, \bL_{\prec j\,]} ), \,
j=1,\dots,q
\right\}$,  where
\be
\bL_{ \prec j\,]}=-\bSigma_{ \prec j\, \succ}\bSigma_{ \prec j\,]}, \quad
\sigma_j^2 = \bSigma_{jj\g\pa(j)};
\ee
see also \cite{cao:et:al:2019}.
Accordingly, Equation \eqref{eq:Gaussian:DAG:facorization} can be written as
\ben
\label{eq:Gaussian:DAG:Chol:facorization}
f(x_1,\dots,x_q\g \bD, \bL)
=
\prod_{j=1}^{q}d\,\N(x_j\g -\bL_{ \prec j\,]}^\top\bx_{\pa(j)}, \sigma_j^2),
\een
with the understanding that the conditional expectation of $X_j$ in \eqref{eq:Gaussian:DAG:Chol:facorization} is zero whenever ${\pa(j)}$ is the empty set.

\black

\subsection{DAG-probit models}
\label{subsec:DAG:probit:models}

We introduce in this section the general form of a \textit{DAG-probit model}. 
We assume that the joint distribution of
$(X_1, X_2, \dots,X_q)$ is Gaussian and Markov w.r.t. $\D$ so that its density is as in \eqref{eq:Gaussian:DAG:Chol:facorization}.
Recall that $X_1$ is latent and we are only allowed to observe the binary variable $Y\in\{0,1\}$.
Specifically, for a given  threshold $\theta_0 \in (-\infty, +\infty)$, we assume that
\ben
\label{eq:latent}
Y =
\begin{cases}
	\,\, 1 & \text{if $X_1 \in [\,\theta_0,+\infty)$},\\
	\,\, 0 & \text{if $X_1 \in (-\infty,\theta_0)$}.
\end{cases}
\een
%
%
%
%
Combining
\eqref{eq:Gaussian:DAG:Chol:facorization} with
\eqref{eq:latent},
the joint density of $(Y,X_1,\dots,X_q)$ becomes
\ben
\label{eq:density:DAG:probit}
\begin{aligned}
	f(y,x_1,\dots,x_q\g \bD,\bL, \theta_0)
	&=f(x_1,\dots,x_q\g \bD, \bL)
	\cdot
	\mathbbm{1}(\theta_{y-1} < x_1 \le \theta_{y}) \\
	&=\left\{
	\prod_{j=1}^{q}
	d\,\N(x_j\g -\bL_{ \prec j\,]}^\top\bx_{\pa(j)}, \sigma_j^2)
	\right\}
	\cdot
	\mathbbm{1}(\theta_{y-1} < x_1 \le \theta_{y}),
\end{aligned}
\een
where $\theta_{-1}=-\infty, \theta_1=+\infty$.
Equation \eqref{eq:density:DAG:probit} defines  a (Gaussian) \textit{DAG-probit model}. A related expression  appears  in \cite{Guo:Levina:et:al:2015} who model a multivariate distribution of ordered  categorical variables through a collection of Gaussian random variables Markov with respect to an undirected graphical model.
Now recall from \eqref{eq:Gaussian:DAG:Chol:facorization} that the
conditional distribution of the latent variable $X_1$ is
$\N(-\bL_{ \prec 1\,]}^\top \bx_{\pa(1)}, \sigma_1^2)$
and, as in standard probit regression, we set $\sigma_1^2=1$  for identifiability reasons.

Finally, by considering $n$ independent samples $(y_i, x_{i,2},\dots,x_{i,q})$, $i=1,\dots,n$,
from
\eqref{eq:density:DAG:probit},
the \emph{augmented} likelihood can be written as
\ben
\label{eq:likelihood:DAG:probit}
\begin{aligned}
f(\by,\bX \g \bD, \bL, \theta_0)
&=
\prod_{i=1}^{n}
f(x_{i,1},\dots,x_{i,q}\g \bD, \bL)
\cdot
\mathbbm{1}(\theta_{y_i-1} < x_{i,1} \le \theta_{y_i}) \\
&=
\prod_{j=1}^{q}
d\,\N_n(\bX_j\g -\bX_{\pa(j)} \bL_{ \prec j\,]}, \sigma_j^2 \bI_n)
\cdot
\left\{
\prod_{i=1}^{n}
\mathbbm{1}(\theta_{y_i-1} < x_{i,1} \le \theta_{y_i})
\right\},
\end{aligned}
\een
where $\by=(y_1\dots,y_n)^\top$,   $\bX$ is the $(n,q)$ augmented data matrix, and $\bX_{A}$
is the submatrix of $\bX$ corresponding to the set $A$ of columns of $\bX$.

\section{Causal effects}
\label{sec:causal:effects}

Consider the joint density  of the random vector $\left(X_1,\dots,X_q\right)$  Markov  w.r.t. a DAG which factorizes as
in \eqref{eq:DAG:facorization};
%
the latter is referred to as the \textit{observational} (or \textit{pre-intervention}) distribution.

We now introduce the notion of \emph{intervention}. A deterministic intervention on variable $X_s$, $s\in\{2,\dots,q\}$ is denoted by $\textnormal{do}(X_s=\tilde{x})$ and consists in setting $X_s$ to the value $\tilde{x}$.
The \textit{post-intervention} density of $(X_1,\dots,X_q)$ is then obtained using the truncated factorization
\ben
\label{eq:intervention:distrib}
f(x_1,\dots,x_{q}\g \textnormal{do}(X_s=\tilde{x}))=
\begin{cases}
	\prod\limits_{j=1,j\ne s}^{q}f(x_j\g \bx_{\pa(j)})|_{x_s=\tilde{x}}
	& \text{if $x_s=\tilde{x}$},\\
	\quad \, 0
	& \text{otherwise},
\end{cases}
\een
where, importantly, each term $f(x_j\g\cdot)$ in \eqref{eq:intervention:distrib} is the corresponding (pre-intervention) conditional distribution of Equation
\eqref{eq:DAG:facorization};
see \citet{Pear:2000}.
We emphasize that the post-intervention density
$f(x_1,\dots,x_{q}\g \textnormal{do}(X_s=\tilde{x}))$
is conceptually distinct from the usual \textit{conditional} density
$f(x_1,\dots,x_{q}\g X_s=\tilde{x})$, which arises out of passive observation of
$X_s=\tilde{x}$.
An important feature of Equation \eqref{eq:intervention:distrib} is that the data generating system is \lq \lq stable\rq \rq{} under exogenous interventions, in the sense that only the local component distribution $f(x_s\g \bx_{\pa(s)})$ is affected by the intervention and effectively reduces to a point mass on $\tilde{x}$. All the remaining terms are  immune to the intervention and thus remain the same.
The post-intervention distribution of the (latent) response $X_1$ is then obtained by integrating \eqref{eq:intervention:distrib} w.r.t. $x_2,\dots,x_q$ which simplifies to
\ben
\label{eq:intervention:distrib:Y}
f(x_1\g \textnormal{do}(X_s=\tilde{x})) =
\int f(x_1\g \tilde{x},\bx_{\pa(s)}) f(\bx_{\pa(s)}) \, d \bx_{\pa(s)};
\een
see also \citet[Theorem 3.2.2]{Pear:2000}.

We now move back to the Gaussian setting of Section \ref{subsec:Gaussian:DAG:models}, and assume that $(X_1,X_2,$ $\dots,X_q) \g \bSigma \sim \N_q(\bzero, \bSigma)$,
where  the covariance matrix $\bSigma$ is
Markov w.r.t. to the underlying DAG.  
The post-intervention distribution of $X_1$ can thus be written as
\ben
\begin{aligned}
\label{eq:intervention:distrib:Y:gaussian}
f(x_1\g \textnormal{do}(X_s= \tilde x),\bSigma) = \int f(x_1\g \tilde x,\bx_{\pa(s)},\bSigma) \cdot f(\bx_{\pa(s)}\g\bSigma) \, d\bx_{\pa(s)} \\
= \int d\,\N(x_1\g\gamma_s \tilde x+\bgamma^\top\bx_{\pa(s)},\delta_1^2)
\cdot d\,\N(\bx_{\pa(s)}\g\bzero,\bSigma_{\pa(s),\pa(s)}) \, d\bx_{\pa(s)},
\end{aligned}
\een
where $\delta_1^2 = \var(X_1\g X_s=\tilde{x},\bx_{\pa(s)}, \bSigma)$.
The following Proposition
gives the analytic form of the post-intervention distribution of $X_1$.
\begin{prop}
	\label{proposition}
	Let $(X_1,X_2,\dots,X_q) \g \bSigma \sim \N_q(\bzero, \bSigma)$ and consider the \textnormal{do} operator $\textnormal{do}(X_s=\tilde{x})$, $s\in\{2,\dots,q\}$. Then the post-intervention distribution of $X_1$ is
	\be
	f(x_1\g \textnormal{do}(X_s=\tilde{x}), \bSigma) = d\,\N\left(x_1\g\gamma_s \tilde x, \frac{\delta_1^2}{1-(\bgamma^\top\bT^{-1}\bgamma)/\delta_1^2}\right),
	\ee
	where
	\begin{eqnarray}
	\delta_1^2 &=& \bSigma_{1\g\fa(s)}, \nonumber \\
	(\gamma_s,\bgamma^\top)^\top &=& \bSigma_{1,\fa(s)}\left(\bSigma_{\fa(s),\fa(s)}\right)^{-1}, \nonumber  \\
	\bT &=& \left(\bSigma_{\pa(s),\pa(s)}\right)^{-1} + \frac{1}{\delta_1^2}\bgamma\bgamma^\top, \nonumber
	\end{eqnarray}
	with the understanding that node $s$ occupies the first position in the set $\fa(s)$.
	\begin{proof}
		See Supplementary material \citep{caste:cons:suppl:BA}.
	\end{proof}
	
\end{prop}

The previous reasoning considered the intervention distribution of the latent response variable $X_1$ following
$\textnormal{do}(X_s=\tilde{x})$.
Typically distribution \eqref{eq:intervention:distrib:Y} is summarized by its  expected value $\vat (X_1 \g \textnormal{do}(X_s=\tilde{x}))$. When $X_s$ is continuous,  one  can
define  the (total) causal effect
 as the derivative
of $\vat (X_1 \g \textnormal{do}(X_s={x}))$ w.r.t. $x$ evaluated at $\tilde{x}$:
this is especially convenient when the expectation is linear, as in the Gaussian case \eqref{eq:intervention:distrib:Y:gaussian}, because  the causal effect admits a simple interpretation: it corresponds to the \lq \lq regression parameter\rq \rq{} $\gamma_s$ associated to variable $X_s$
\citep{Maat:etal:2009}.
Our interest however lies in  the observable response  variable $Y$, and therefore we aim to evaluate
$\vat(Y\g\textnormal{do}(X_s=\tilde{x}),\bSigma, \theta_0)$.
We thus obtain
\ben
\label{eq:causal:effect}
\begin{aligned}
\vat(Y\g\textnormal{do}(X_s=\tilde{x}),\bSigma, \theta_0)
&=
\textnormal{Pr}(Y=1\g\textnormal{do}(X_s=\tilde{x}),\bSigma, \theta_0) \\
&=
\textnormal{Pr}(X_1\ge \theta_0 \g \textnormal{do}(X_s=\tilde{x}), \bSigma) \\
&=
1-\Phi\left(\frac{\theta_0-\gamma_s\tilde{x}}{\tau_1}\right)
\equiv \beta_s(\tilde x,\bSigma, \theta_0),
\end{aligned}
\een
where $\Phi(\cdot)$ denotes the c.d.f. of a standard normal and $\tau^2_1=\delta_1^2/\left(1-(\bgamma^\top\bT^{-1}\bgamma)/\delta_1^2\right)$.
One could then compute the partial derivative of
$\vat(Y\g\textnormal{do}(X_s=\tilde{x}),\bSigma, \theta_0)$  w.r.t $x$ evaluated at $\tilde x$,
and obtain
 $
 \phi(
 {\theta_0-\gamma_s\tilde{x}}/{\tau_1})
 {\gamma_s}/{\tau_1},
 $
where $\phi(\cdot)$ is the density function of a standard normal. This however would still depend on $\tilde{x}$.
For this reason, and because
\eqref{eq:causal:effect} enjoys an intuitive interpretation being a probability, we will simply denote
$\textnormal{Pr}(Y=1\g\textnormal{do}(X_s=\tilde{x}),\bSigma, \theta_0)$ at the causal effect on $Y$ due to an intervention
$\textnormal{do}(X_s=\tilde{x})$.
Finally, we remark that the causal effect of
$\textnormal{do}(X_s=\tilde{x})$ on $Y$, besides being
 a function of the value $\tilde{x}$, depends on $\theta_0$ as well as on the covariance matrix $\bSigma$, where the latter is constrained to be Markov w.r.t. the underlying DAG.

\section{Bayesian inference}
\label{sec:bayesian:inference}



In this section we introduce priors for $(\bOmega,\theta_0, \D)$, which we structure as $p(\bOmega, \theta_0, \D)=p(\bOmega \g \D)p(\D)p(\theta_0)$.
Further distributional results useful for our MCMC scheme of Section \ref{sec:MCMC} are also presented.
\black
We briefly preview here the essential features.

To start with, consider  $p(\bOmega \g \D)$,  $\bOmega \in  \mathcal{P}_{\D}$.
We  first proceed to the re-parameterization $\bOmega \mapsto (\bD,\bL)$ presented in Subsection \ref{subsec:Gaussian:DAG:models}, and
specify a DAG-Wishart prior \citep{cao:et:al:2019} on the  Cholesky parameters $(\bD,\bL)$. We achieve this goal using a highly parsimonious elicitation procedure, which we detail in Section \ref{subsec:prior:omega}.
%
\black
For the unknown threshold $\theta_0\in(-\infty,+\infty)$, we assume a uniform prior, so that $p(\theta_0)\propto 1$ (Section \ref{subsec:prior:theta0}).
Finally, a prior on DAG $\D$ is assigned through independent  Bernoulli distributions on the elements of the skeleton of $\D$ (Section \ref{subse:prior:dag}).

\subsection{Prior on the Cholesky parameters}

\label{subsec:prior:omega}

Consider first a DAG $\D=(V,E)$ which is \emph{complete}, so that the precision matrix $\bOmega$ is unconstrained. A standard conjugate prior is  the Wishart distribution, $\bOmega \sim \W_q(a,\bU)$ having expectation $a \bU^{-1}$,
where $a>q-1$ and $\bU$ is a s.p.d. matrix.
The induced prior on the Cholesky parameters consistent with the DAG parent ordering is such that the node parameters $(\sigma_j^2, \bL_{\prec j\,]}) $, $j=1, \ldots, q$, are independent with distribution
\ben
\label{eq:prior-sigma-L-from-Wishart}
\sigma_j^2 &\sim& \textnormal{I-Ga}\left(\frac{a_j}{2}-\frac{|\pa(j)|}{2}-1,
\frac{1}{2}\bU_{jj|\prec j\succ}\right), \nonumber \\
\bL_{\prec j\,]}\g\sigma_j^2 &\sim& \N_{|\pa(j)|}\left(-\bU_{\prec j \succ}^{-1}\bU_{\prec j\,]},\sigma_j^2\bU_{\prec j \succ}^{-1}\right),
\een
where $|A|$ is the number of elements in the set $A$,
$a_j=a+q-2j+3$; see \citet[Supplemental B]{Ben:Massam:arXiv}.
The symbol $\textnormal{I-Ga}(a,b)$ stands for an Inverse-Gamma distribution with shape  $a>0$ and rate $b>0$ having expectation $b/(a-1)$ ($a>1$).
In absence of substantive prior information,  a standard choice for the hyperparameter $\bU$,
hereinafter adopted,  is $\bU=g\,\bI_q$, where $g>0$ and $\bI_q$ is the $(q,q)$ identity matrix. It is straightforward  to show that \eqref{eq:prior-sigma-L-from-Wishart} reduces to
\ben
\label{eq:prior:chol:complete}
\sigma_j^2 &\sim& \textnormal{I-Ga}\left(\frac{a_j}{2}-\frac{|\pa(j)|}{2}-1,
\frac{1}{2}\,g\right), \nonumber \\
\bL_{\prec j\,]}\g\sigma_j^2 &\sim& \N_{|\pa(j)|}\left(\bzero,\frac{1}{g}\sigma_j^2\,\bI_{|\pa(j)|}\right).
\een
In addition, to guarantee the identifiability of the DAG-probit model,  we fix $\sigma_{1}^2=1$, so that instead of  $p(\sigma_{1}^2, \bL_{\prec 1\,]} )$ we need only to consider  $p(\bL_{\prec 1\,]})$ with
$\bL_{\prec 1\,]} \sim \N_{|\pa(1)|}(\bzero,$ $g^{-1}\,\bI_{|\pa(1)|})$; see also Section \ref{subsec:DAG:probit:models}.
Recall that  \eqref{eq:prior:chol:complete} applies only to a complete DAG $\D$.

Consider now the case in which $\D$ is not complete. The idea is to leverage \eqref{eq:prior:chol:complete} to construct a prior  on $(\bD, \bL)$, the Cholesky parameters of $\bOmega \in \mathcal{P}_{\D}$, which is valid for \textit{any} $\D$.
To this end we follow the procedure of
\citet{Geig:Heck:2002}.
Specifically, let $\D$ be an arbitrary DAG and assume a parent ordering of its nodes. For each node $j \in \{1,\dots,q\}$, let  $\big\{\sigma_j^2,\bL_{\prec j \, ]}\big\}$ be the Cholesky parameters associated to node  $j$, and identify a \textit{complete} DAG $\D^{C(j)}$ such that $\pa_{\D^{C(j)}}(j')=\pa_{\D}(j)$, where $j'$ in $\D^C(j)$ corresponds to the same variable as $j$ in $\D$. Because of the parent ordering $j'=q-|\pa_\D(j)|$ which is usually different from $j$.
Let
$\big\{\sigma_{j'}^{2 \,C(j)},\bL_{\prec j' \, ]}^{C(j)}\big\}$ be the Cholesky parameters of node $j'$ under the complete DAG $\D^{C(j)}$.
We then assign to $\left\{\sigma_j^2, \bL_{\prec j \, ]}\right\}$ the same prior of $\big\{\sigma_{j'}^{2\,{C(j)}},\bL_{\prec j' \, ]}^{C(j)}\big\}$ which can be gathered from Equation \eqref{eq:prior:chol:complete} in the complete DAG-Wishart version.
In particular $\sigma_j^2\sim \textnormal{I-Ga}((a+|\pa_{\D}(j)|-q+3)/2-1,g/2)$ and $\bL_{\prec j\,]}$ is distributed as a zero-mean multivariate normal with covariance matrix $g^{-1}\sigma_j^2\bI_{|\pa_\D(j)|}$. Therefore  both distributions only depend on the cardinality of $\pa_\D(j)$ which is the same across alternative parent orderings.
Finally, by assuming independence among node-parameters $(\sigma_j^2, \bL_{\prec j\,]})$, we can write
\ben
\label{eq:prior:fact}
p(\bD, \bL)=\prod_{j=1}^{q}p(\sigma_j^2, \bL_{\prec j\,]}), \quad (\bL, \bD) \in {\Theta}_{\D},
\een
where ${\Theta}_{\D}$ is the   image of the space $\mathcal{P}_{\D}$ under the mapping $\bOmega \mapsto (\bD,\bL)$.

\subsection{Prior on DAG space} 
\label{subse:prior:dag}

For a given DAG  $\D$,
let $\bA^{\D}$ be the (symmetric) 0-1 adjacency matrix of the skeleton of $\D$ whose $(u,v)$ element is denoted by $\bA^{\D}_{(u,v)}$.
Conditionally on the edge inclusion  probability $\pi$,
we first assign a Bernoulli prior independently to each element $\bA^{\D}_{(u,v)}$ belonging to the lower-triangular part, that is:
$\bA^{\D}_{(u,v)} \g \pi \, \stackrel{iid}{\sim} \, \textnormal{Ber}(\pi),  u > v$.
As a consequence we get
\ben
p(\bA^{\D})=
\pi^{|\bA^{\D}|}(1-\pi)^{\frac{q(q-1)}{2}-|\bA^{\D}|},
\een
where $|\bA^{\D}|$ denotes the number of edges in the skeleton, equivalently the number of entries equal to one in the lower-triangular part of $\bA^{\D}$.
Finally, we set
$p(\D)\propto p(\bA^{\D})$,
for $\D \in \mathcal{S}_q$, where $\mathcal{S}_q$ is the set of all DAGs on $q$ nodes.

\subsection{Posterior distribution of $\theta_0$}
\label{subsec:prior:theta0}

As mentioned, in absence of substantive prior information,
we assign a flat improper prior to the treshold $\theta_0\in(-\infty,\infty)$, $p(\theta_0)\propto 1$.
Accordingly, we need to prove that the posterior of $\theta_0$ is proper. The next proposition details under which conditions 2 is guaranteed.

\begin{prop}
	\label{prop:posterior:theta_0}
	Under the prior \eqref{eq:prior:chol:complete} for $(\bD,\bL)$, $p(\D)$ as in Section \ref{subse:prior:dag} for DAG $\D$ and the improper prior $p(\theta_0)\propto 1$ for $\theta_0$, the posterior of $\theta_0$ is proper provided the sample contains at least two observations with distinct values for $Y$, that is $y_i=1$, $y_{i'}=0$ ($ i \neq i'$).
	\begin{proof}
		See Supplementary material \citep{caste:cons:suppl:BA}.
	\end{proof}
\end{prop}
Additionally, we prove in the Supplementary material that under the conditions of Proposition \ref{prop:posterior:theta_0} the joint posterior of $(\bD,\bL,\D,\theta_0, \bX_1)$ is proper.
Clearly, alternative priors for $\theta_0$ might have been used; yet the full conditional of $\theta_0$ would not be amenable to direct sampling. As a consequence, posterior inference on $\theta_0$ is performed through a Metropolis Hastings step inside our MCMC scheme; see Section \ref{sec:MCMC} for details.

\section{MCMC scheme}
\label{sec:MCMC}
In this section we detail the MCMC scheme that we adopt to target the posterior distribution
%
%
\ben
\label{eq:posterior}
p(\bD, \bL,\D, \theta_0, \bX_1 \g
\by, \bX_{-1})
\propto
f(\by,\bX \g \bD, \bL, \D, \theta_0)
\, p(\bD, \bL \g \D) \, p(\D),
\een
now emphasizing the dependence on DAG $\D$,
where $\bX_{-1}=(\bX_2, \ldots,\bX_q)$, and the term $p(\theta_0)$ has been omitted because it is proportional to one.

\subsection{Update of $(D,L,\D)$}

From \eqref{eq:posterior} the full conditional distribution of $(\bD,\bL,\D)$ is
\be
p(\bD,\bL,\D\g\by,\bX,\theta_0)\propto p(\bX\g \bD,\bL,\D)p(\bD,\bL\g\D)p(\D)
\ee
using \eqref{eq:likelihood:DAG:probit},
where $\bX=(\bX_1,\bX_2,\dots,\bX_q)$ is the $(n,q)$ augmented data matrix.

To sample from $p(\bD,\bL,\D\g\bX)$ we adopt a reversible jump MCMC algorithm which takes into account the partial analytic structure
(PAS, \citealt{Godsill:2012}) of the DAG-Wishart distribution to sample DAG $\D$ and the Cholesky parameters $(\bD,\bL)$
from their full conditional. A similar approach was implemented in \cite{wang:li:2012} for Gaussian undirected graphical models using G-Wishart priors.
Details about the PAS algorithm and its implementation in our DAG setting are reported in the Supplementary material \citep{caste:cons:suppl:BA}.

Specifically, at each iteration of the MCMC scheme, we first propose a new DAG $\D'$ from a suitable proposal distribution $q(\D'\g\D)$; see again our Supplementary material.
In particular, it is shown that when proposing a DAG $\D'$
which differs from the current graph $\D$ by one edge $(h,j)$,
the acceptance probability for $\D'$ is  given by
\ben
\label{eq:ratio:PAS:reduced}
\alpha_{\D'}=\min
\left\{
1,\frac{m_{}(\bX_j\g\bX_{\pa_{\D'}(j)}, \D')}
{m_{}(\bX_j\g\bX_{\pa_{\D}(j)}, \D)}
\cdot\frac{p(\D')}{p(\D)}
\cdot\frac{q(\D\g\D')}{q(\D'\g\D)}
\right\}
\een
where, for $j\in\{2,\dots,q\}$,
\ben
\label{eq:marg:like:j}
m_{}(\bX_j\g\bX_{\pa_{\D}(j)}, \D) =
(2\pi)^{-\frac{n}{2}}
\frac{\big|\bT_j\big|^{1/2}}
{\big|\bar{\bT}_j\big|^{1/2}}\cdot
\frac{\Gamma\left(a_j^*+\frac{n}{2}\right)}{\Gamma\left(a_j^*\right)}
\left[\frac{1}{2}g\right]^{a_j^*}
\left[\frac{1}{2}\big(g+\bX_j^\top\bX_j - \hat{\bL}_j^\top\bar{\bT}_j\hat{\bL}_j\big)\right]^{-(a_j^*+n/2)}
\een
with
\be
\bT_j &=& g\bI_{|\pa_{\D}(j)|} \\
\bar{\bT}_j &=& g\bI_{|\pa_{\D}(j)|}+\bX_{\pa_{\D}(j)}^\top\bX_{\pa_{\D}(j)} \\
\hat{\bL}_j
&=& \big(g\bI_{|\pa_{\D}(j)|}+\bX_{\pa_{\D}(j)}^\top\bX_{\pa_{\D}(j)}\big)^{-1}\bX_{\pa_{\D}(j)}^{\top}\bX_j,
\ee
$a_j^*=\frac{a_j}{2}-\frac{|\pa_{\D}(j)|}{2}-1$ and $\bX_{\pa_{\D}(j)}$ denotes the $(n,|\pa_{\D}(j)|)$ sub-matrix of $\bX$ whose columns belong to the set
$\pa_{\D}(j)$. For $j=1$, because we fixed $\sigma_1^2=1$, we have instead
\ben
\label{eq:marg:like:1}
m_{}(\bX_1\g\bX_{\pa_{\D}(1)}, \D) =
(2\pi)^{-\frac{n}{2}} \frac{\big|\bT_1\big|^{1/2}}
{\big|\bar{\bT}_1\big|^{1/2}}\cdot
\exp\left\{-\frac{1}{2}\big(\bX_1^\top\bX_1+
\hat{\bL}_1^\top\bar{\bT}_1\hat{\bL}_1\big)\right\},
\een
with $\bT_1,\bar{\bT}_1,\hat{\bL}_1$ defined  in analogy with the expressions appearing after \eqref{eq:marg:like:j}; see the Supplementary material \citep{caste:cons:suppl:BA} for details.
Moreover, given DAG $\D$ and $\bX_1$,
the full conditional of $(\bD,\bL)$ reduces to
the augmented posterior $p(\bD,\bL\g\bX)$,
which is conditional on the actual data $(\bX_2, \ldots, \bX_q)$ as well as the latent values $\bX_1$ and can be easily sampled from.
Specifically, since
%
%
%
%
\ben
f(\bX \g \bD,\bL)=
\prod_{j=1}^{q}
d\,\N_n(\bX_j\g -\bX_{\pa_{\D}(j)} \bL_{ \prec j\,]}, \sigma_j^2 \bI_n)
\een
and because of \eqref{eq:prior:fact}
and conjugacy of the Normal-Inverse-Gamma prior in \eqref{eq:prior:chol:complete} with the Normal density, the posterior distribution of the Cholesky parameters  given  $\bX$
is, for $j=2,\dots,q$,
%
\ben
\label{eq:posterior:L:D}
\sigma_j^2\g\bX &\sim& \textnormal{I-Ga}\left(a_j^*+\frac{n}{2},
\frac{1}{2}\big(g+\bX_j^\top\bX_j - \hat{\bL}_j^\top\bar{\bT}_j\hat{\bL}_j\big)\right), \nonumber \\
\bL_{ \prec j\,]}\g\sigma_j^2,\bX &\sim& \N_{|\pa_{\D}(j)|}
\big(-\hat{\bL}_j,
\sigma_j^2\bar{\bT}_j^{-1}\big).
\een
Moreover, for node $1$ where $\sigma_{1}^2=1$, we have
\ben
\bL_{ \prec 1\,]}\g\bX &\sim& \N_{|\pa_{\D}(1)|}
\big(-\hat{\bL}_1,
\bar{\bT}_1^{-1}\big).
\een

\black


\black

\subsection{Update of $X_1$ and $\theta_0$}

Updating of $\bX_1=(x_{1,1},\dots,x_{n,1})^\top$ can be performed by direct sampling from the full conditional distribution of each latent observation $x_{i,1}$,
\be
f(x_{i,1}\g y_i,x_{i,2},\dots,x_{i,q},\bD, \bL,\D, \theta_0)
\propto
f(x_{i,1}\g\bx_{i,\pa_{\D}(1)},\bL_{\prec j\,]})
\cdot
\mathbbm{1}(\theta_{y_i-1} < x_{i,1} \le \theta_{y_i}),
\ee
which corresponds to a $\N(-\bL_{\prec 1\,]}^\top \bx_{i,\pa_{\D}(1)}, 1)$ truncated at the interval $(\theta_{y_i-1},\theta_{y_i}]$.

Finally, the cut-off $\theta_0$ is updated through a Metropolis Hastings step where, given the current value $\theta_0$, a candidate value $g_0$ is proposed from $q(g_0\g\theta_0)=d\,\N(g_0\g\theta_0,\sigma_0^2)$. We then set $\theta_0=g_0$ with probability
\ben
\label{eq:accept:theta:0}
\alpha_\theta=\min\left\{1;r_\theta\right\},
\een
where
\be
r_\theta\,
=\,
\frac{\prod\limits_{i=1}^n
	\left[\Phi\big(g_{y_i}\g-\bL_{\prec 1\,]}^{\top}\bx_{i,\pa_{\D}(1)},1\big)-
	\Phi\big(g_{y_i-1}\g-\bL_{\prec 1\,]}^{\top}\bx_{i,\pa_{\D}(1)},1\big)
	\right]}
{\prod\limits_{i=1}^n
	\left[\Phi\big(\theta_{y_i}\g-\bL_{\prec 1\,]}^{\top}\bx_{i,\pa_{\D}(1)},1\big)-
	\Phi\big(\theta_{y_i-1}\g-\bL_{\prec 1\,]}^{\top}\bx_{i,\pa_{\D}(1)},1\big)
	\right]}
\cdot
\frac{d\,\N\left(\theta_0\g g_0,\sigma_0^2\right)}
{d\,\N\left(g_0\g \theta_0,\sigma_0^2\right)},
\ee
and $g_{-1}=\infty, g_{1}=+\infty$.

\subsection{Algorithm}

Algorithm \ref{alg:MCMC} summarizes our MCMC scheme. The output is a collection of DAGs $\big\{\D^{(t)}\big\}_{t=1}^{T}$ and Cholesky parameters $\big\{\big(\bD^{\D^{(t)}},\bL^{\D^{(t)}}\big)\big\}_{t=1}^{T}$
approximatively sampled from the target distribution \eqref{eq:posterior}.
In particular we can compute posterior summaries of interest such as the posterior probabilities of edge inclusion, namely
\ben
\label{eq:posterior:edge:inclusion}
\hat p_{u\rightarrow v}(\by, \bX_2,\dots,\bX_q)\equiv \hat p_{u\rightarrow v}=\frac{1}{T}\sum_{t=1}^{T}\mathbbm{1}_{u\rightarrow v}\big\{\D^{(t)}\big\},
\een
where $\mathbbm{1}_{u\rightarrow v}\big\{\D^{(t)}\big\}$  takes value 1 if and only if $\D^{(t)}$ contains the edge $u\rightarrow v$.
Moreover, we can reconstruct the covariance matrices $\big\{\bSigma^{\D^{(t)}}\big\}_{t=1}^{T}$ using \eqref{eq:cholesky:fact}.
The latter can be subsequently retrieved to obtain for selected $s\in\{2,\dots,q\}$ and intervention value $\tilde{x}$
the collection of causal effects $\big\{\beta_s^{(t)}(\tilde{x})\big\}_{t=1}^{T}$
defined in \eqref{eq:causal:effect}, where we set $\beta_s^{(t)}(\tilde{x}) \equiv \beta_s\big(\tilde{x},\bSigma^{\D^{(t)}},\theta_0^{(t)}\big)$.
An overall summary of the causal effect of $\textnormal{do}(X_s=\tilde{x})$ on $Y$ can be computed as
\ben
\label{eq:causal:effect:BMA}
\hat{\beta}_s^{BMA}(\tilde{x})=\frac{1}{T}\sum_{t=1}^{T}\beta_{s}^{(t)}(\tilde{x}),
\een
which corresponds to a Bayesian Model Averaging (BMA) estimate where posterior (DAG) model probabilities are approximated through the MCMC frequencies of visits; see  \cite{Garcia:Donato:et:al:2013} for a discussion of the merits of  frequency based estimators in large model spaces.

\begin{algorithm}{
		\SetAlgoLined
		\vspace{0.1cm}
		\KwInput{A dataset $(\by,\bX_2,\dots,\bX_q)$ \\
			\KwOutput{$T$ samples from the posterior \eqref{eq:posterior}}
			Initialize $\D^{(0)}$, e.g. the empty DAG, the cut-offs $\theta_{-1}^{(0)}=-\infty,\theta_{0}^{(0)}=0,\theta_1^{(0)}=+\infty$ and the latent variables $\bx_1^{(0)}$, e.g. $x_{i,1}^{(0)}\stackrel{ind}\sim\N(0,1)$ truncated at $(\theta^{(0)}_{y_i-1},\theta^{(0)}_{y_i}]$\;
			\For{$t=1,\dots,T$}{
				Sample $\D'$ from $q(\D'\g\D^{(t-1)})$ and set $\D^{(t)}=\D'$ with probability $\alpha_{\D}$ \eqref{eq:ratio:PAS:reduced}, otherwise $\D^{(t)}=\D^{(t-1)}$\;
				Sample $\big(\bD^{\D^{(t)}}, \bL^{\D^{(t)}}\big)$ from its augmented posterior distribution \eqref{eq:posterior:L:D}\;
				For $i=1,\dots,n$, independently sample $x_{i,1}$ from
				$\N\big(-\bL_{\prec 1\,]}^{\D^{(t)}\top}\bx_{i,\pa_{\D^{(t)}}(1)},1\big)$ truncated at $(\theta_{y_i-1}^{(t)},\theta_{y_i}^{(t)}]$\;
				Propose a cut-off $g_0$ from $q(g_0\g\theta_0^{(t)})$ and set $\theta_0^{(t)}=g_0$ with probability $\alpha_\theta$ \eqref{eq:accept:theta:0}, otherwise $\theta_0^{(t)}=\theta_0^{(t-1)}$; set $\theta_{-1}^{(t)}=-\infty,\theta_{1}^{(t)}=+\infty$.
			}
		}
		\caption{MCMC scheme to sample from \eqref{eq:posterior}}
		\label{alg:MCMC}
	}
\end{algorithm}

\section{Simulations}
\label{sec:simulations}

In this section we evaluate the performance
of our method through simulation studies.
Specifically, for each
combination of number of nodes  $q \in
\{20,40\}$ and sample size
$n\in\{100,200,500\}$, which we call
simulation \emph{scenario},
we generate $40$ DAGs using a  probability of
edge inclusion equal to $p=3/(2q-2)$ to
induce sparsity; see \cite{Pete:Buhl:2014}.
For each DAG $\D$ we then proceed as follows.
We identify a parent ordering and
fix $\bD^{\D}=\bI_q$ and then
randomly sample the entries of $\bL^{\D}$ in
the interval  $[-2,-1]\cup[1,2]$;
next we generate a dataset consisting of $n$
i.i.d. $q$-dimensional  observations from the
structural equation model
\eqref{eq:structural:equation:model} which
also includes the $(n,1)$ vector of latent
observations; we finally fix the threshold $\theta_0=0$ and obtain the 0-1 vector of responses $\by=(y_1,\dots,y_n)^\top$ as in \eqref{eq:latent}.

We apply Algorithm \ref{alg:MCMC} to
approximate the target distribution in
\eqref{eq:posterior} by setting  the number
of MCMC iterations $T= 25000$ for $q=20$, and
$T=50000$ for $q=40$. We also set $g=1/n$ and
$a = q+1$ in the prior on the Cholesky
parameters of $\bOmega$
\eqref{eq:prior:chol:complete} and
$\sigma_0^2=0.25$ in the proposal density for
the cut-off $\theta_0$.

\begin{landscape}
	\begin{figure}
		\begin{center}
			\begin{tabular}{ccc}
				$\quad\quad n = 100$ & $\quad\quad n = 200$ & $\quad\quad n = 500$\\
				\vspace{0.1cm} \\
				\includegraphics[scale=0.2]{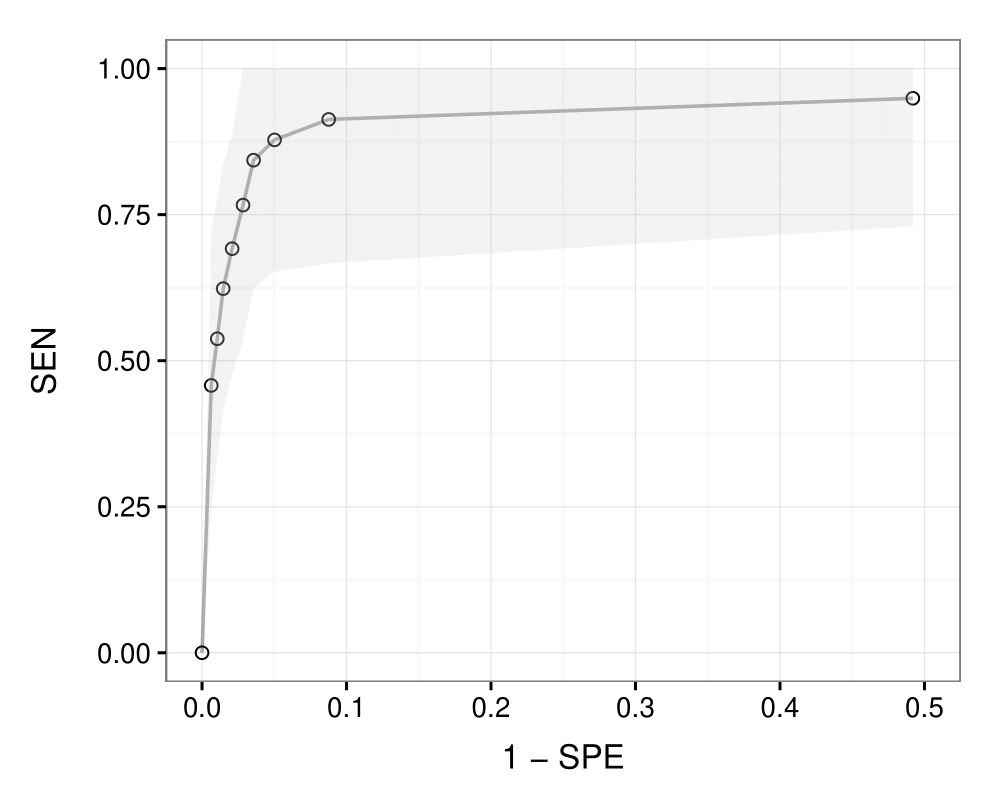}&
				\includegraphics[scale=0.2]{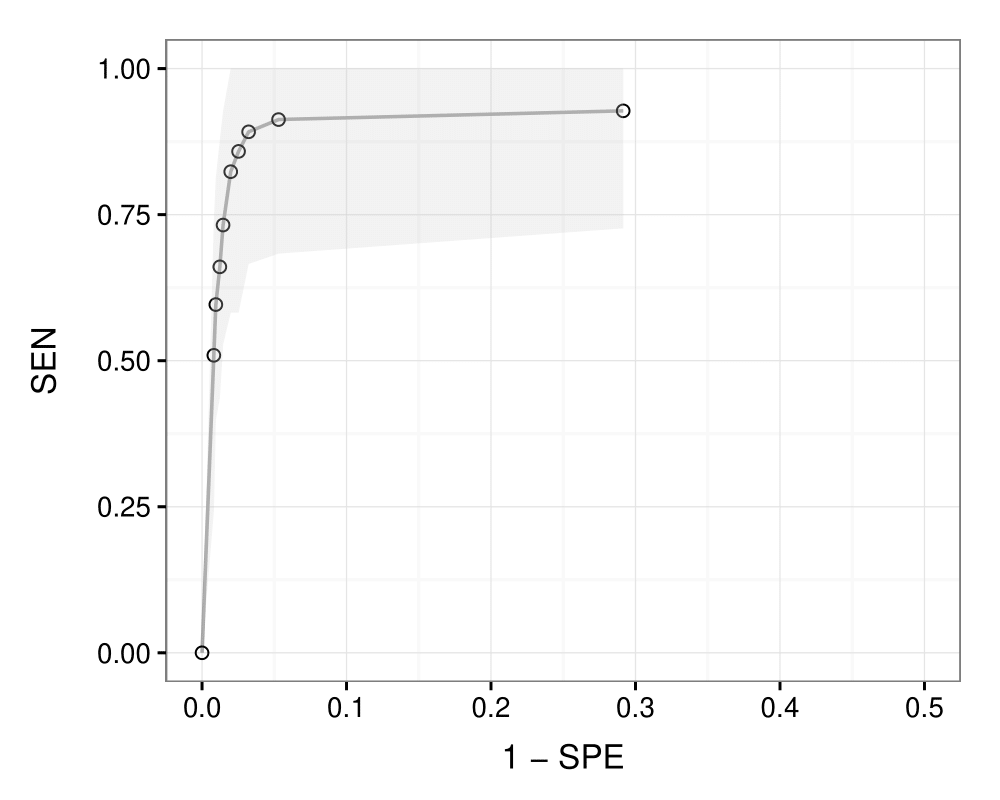}&
				\includegraphics[scale=0.2]{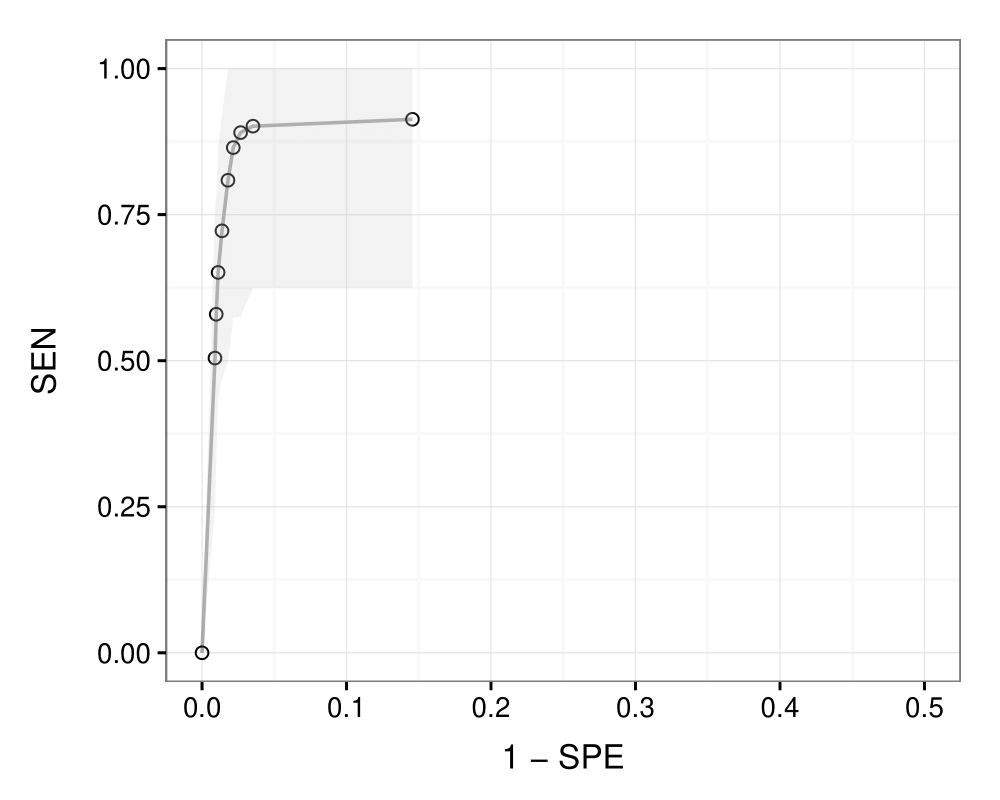}
				\\
				\includegraphics[scale=0.2]{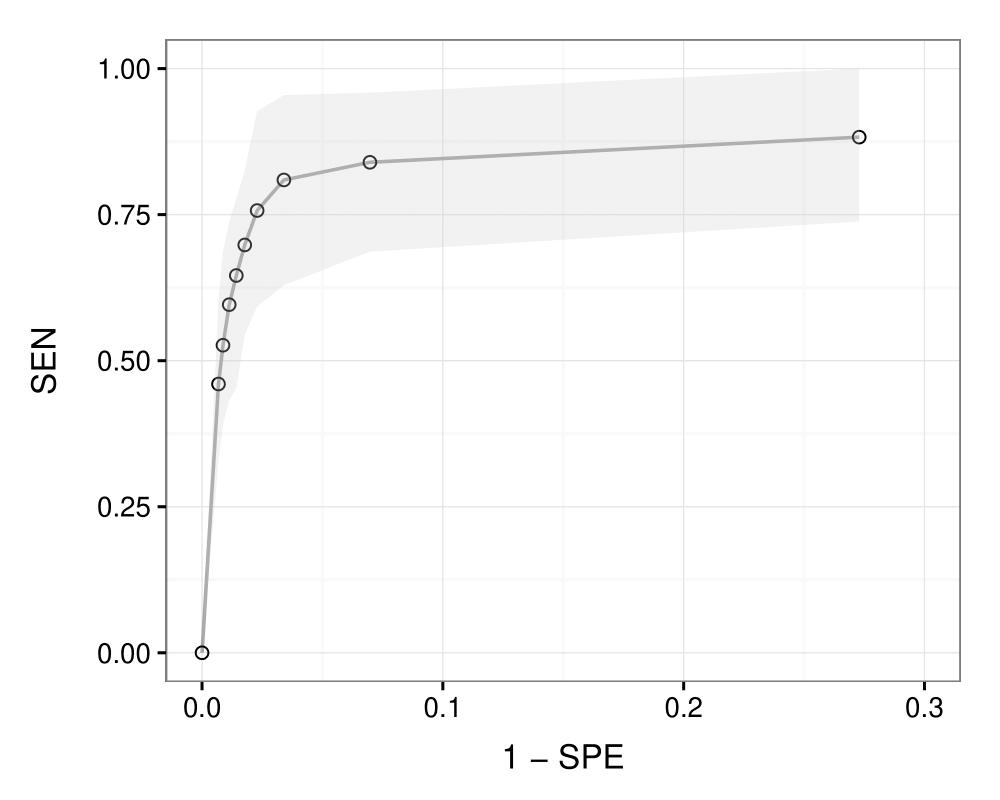}&
				\includegraphics[scale=0.2]{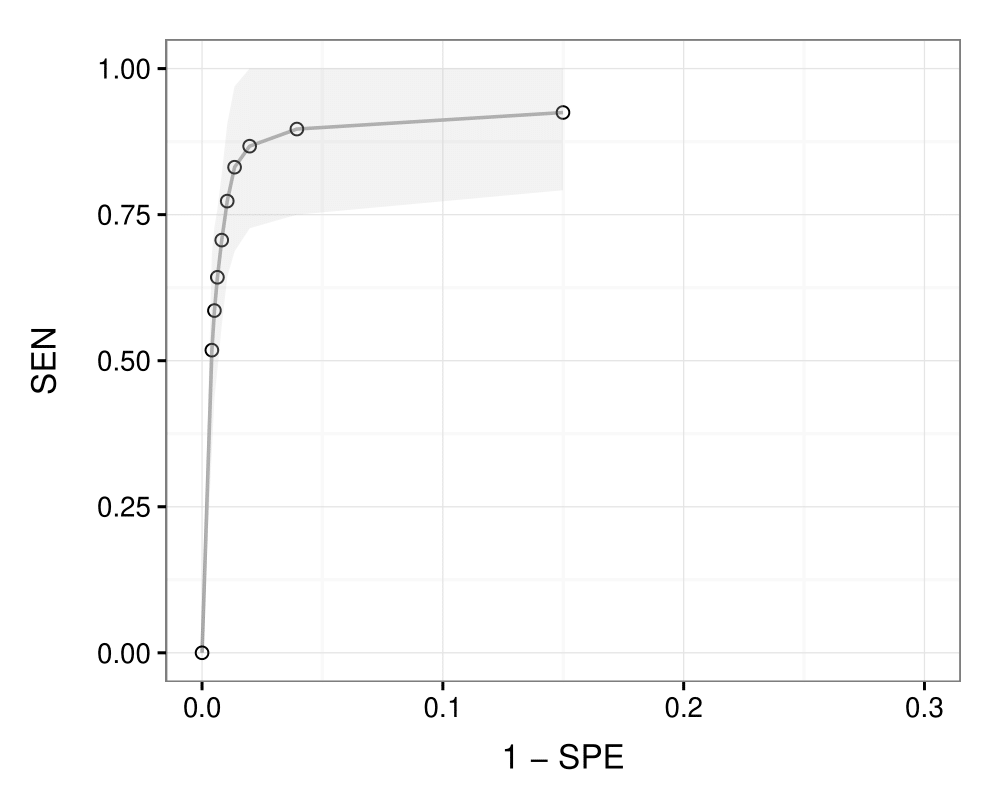}&
				\includegraphics[scale=0.2]{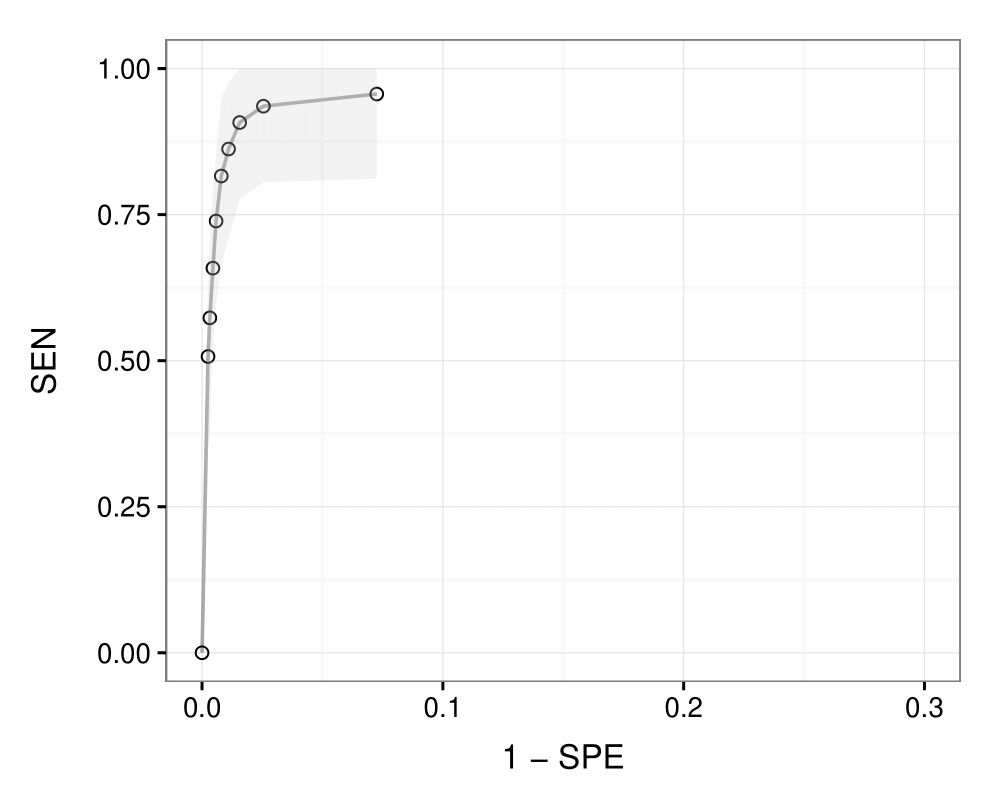}
			\end{tabular}
			\caption{Simulations. Receiver operating characteristic (ROC) curve obtained under varying thresholds for the posterior	probabilities of edge inclusion for each combination of number of nodes $q=\{20,40\}$ (first and second row respectively) and sample size $n\in\{100,200,500\}$. Dots and connecting line describe the (average over the 40 simulated DAGs) ROC curve, while the grey area represents the 5th-95th percentile band.}
			\label{fig:sim:rocs:graph:selection}
		\end{center}
	\end{figure}
\end{landscape}

We begin by evaluating the global performance
of our method in learning the graph
structure.
To this end, we first  estimate the posterior
probabilities of  edge
inclusion by computing $\hat{p}_{u\rightarrow
	v}(\cdot)$ in
\eqref{eq:posterior:edge:inclusion} for each
pair of distinct nodes $u,v$.
Next, we consider a threshold for edge
inclusion $k\in[0,1]$ and for a given $k$
construct a DAG estimate by including those
edges $(u,v)$ whose posterior probability
exceeds $k$.
The resulting graph is compared with the true
DAG through the sensitivity (SEN) and
specificity (SPE) indexes, respectively
defined as
\be
SEN = \frac{TP}{TP+FN}, \quad
SPE = \frac{TN}{TN+FP},
\ee
where $TP, TN, FP, FN$ are the numbers of
true positives, true negatives, false
positives
and false negatives respectively.
The two indexes are used to construct a
receiver operating characteristic (ROC)
curve.
Specifically, for each scenario defined by
$q$ and $n$,  we present a ROC curve
constructed as follows.  For each threshold
$k$,  we compute $SEN$ and $(1-SPE)$ under
each of the 40 DAGs used in the simulation.
The point whose coordinates are the mean
of each of the two measures corresponds to
one  dot
in Figure \ref{fig:sim:rocs:graph:selection}.
The collection of dots
connected by  lines represents an average ROC curve. We proceed similarly to compute the 5th and 95th
percentile and obtain the grey band.
%
%
%
%
%
%
%
%
%

\black

To better appreciate Figure \ref{fig:sim:rocs:graph:selection}, we also compute, for each simulation scenario $(q,n)$, the area under the (average) ROC curve (AUC) whose values  are reported in Table \ref{tab:AUC}.
They are close or above 94\% under the three sample sizes considered when $q=20$. When $q=40$ AUC exceeds 90\% for $n=100$ and rises to over 97\% for $n=500$.

\begin{table}[ht]
	\centering
	\begin{tabular}{rccc}
		\hline
		& $n = 100$ & $n = 200$ & $n = 500$ \\
		\hline
		$q = 20$ & 93.89 & 94.19 & 95.12 \\	
		$q = 40$ & 90.94 & 94.91 & 97.19 \\
		\hline
	\end{tabular}
	\caption{Simulations. Area under the curve (percentage values) computed from the average ROC curves in Figure \ref{fig:sim:rocs:graph:selection} for number of nodes $q \in\{20,40\}$ and sample sizes $n\in\{100,200,500\}$.}
	\label{tab:AUC}
\end{table}

\begin{figure}
	\begin{center}
		\begin{tabular}{cc}
			$\quad\quad q = 20$ & $\quad\quad
			q = 40$ \\
			\vspace{0.05cm} \\
			\includegraphics[scale=0.22]{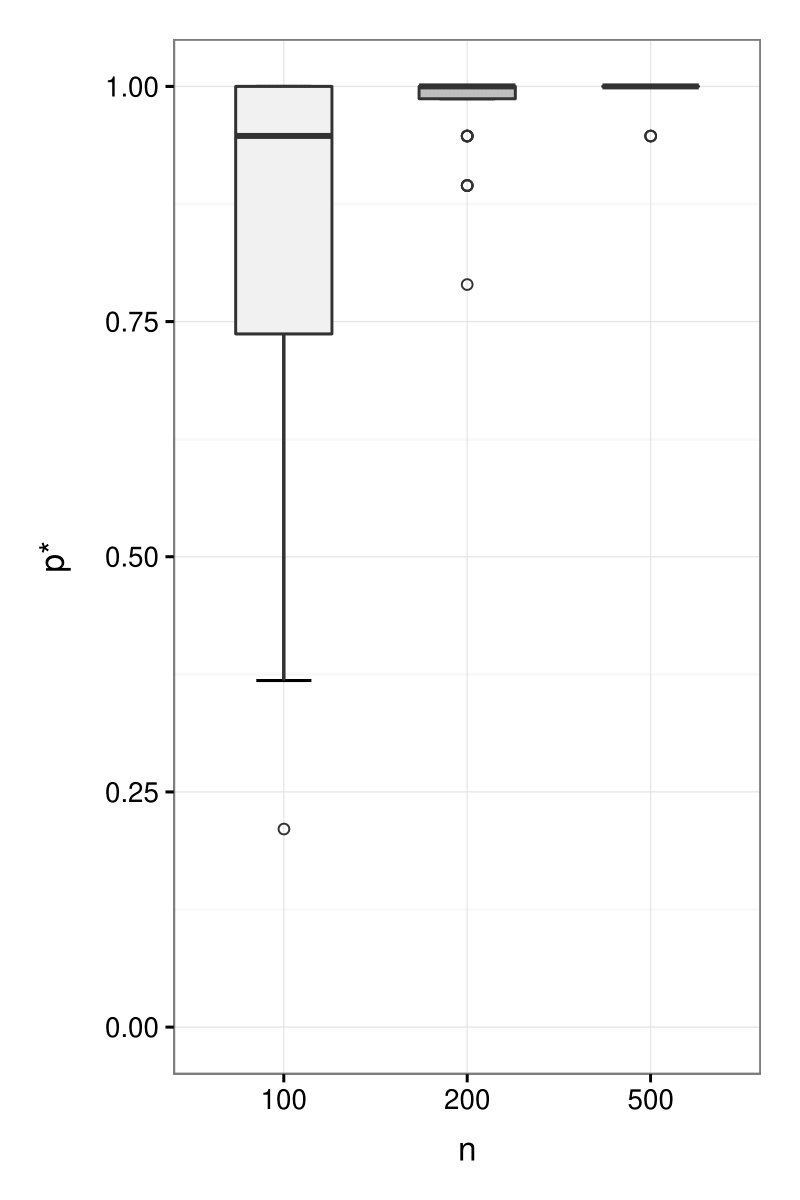}
			&
			\includegraphics[scale=0.22]{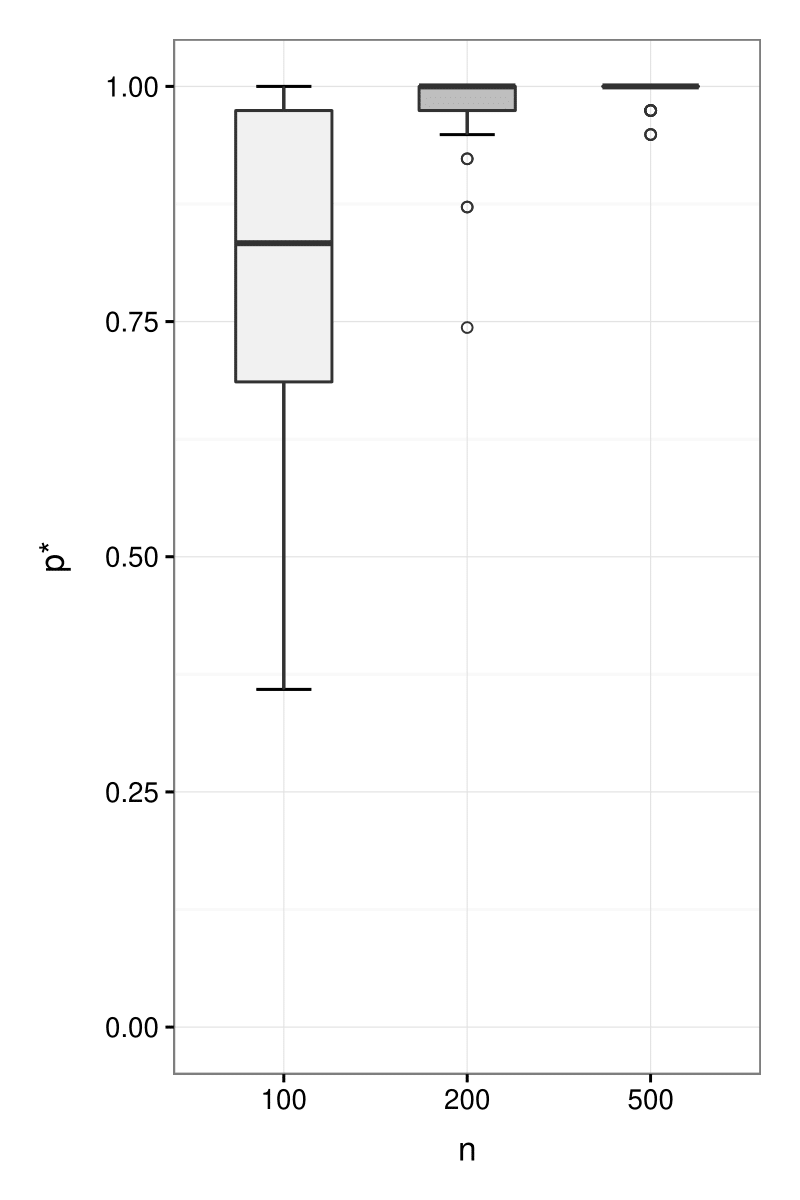}
		\end{tabular}
		\caption{Simulations. Distribution
			across 40 simulated datasets of the
			proportion of predictors $p^*$ that are
			correctly classified given a threshold for edge inclusion $k^*=0.5$ for each combination of number of nodes
			$q\in\{20,40\}$ and sample size
			$n\in\{100,200,500\}$.}
		
		\label{fig:sim:boxplots:covariate:selection}
	\end{center}
\end{figure}

A more  specific check  on the ability of our method in recovering the structure of the underlying DAG can be considered.
Since $Y$ is the response, interest centers on the causal effect on $Y$ following an intervention on a variable in the system.
A natural group of intervention variables is represented by the set of parents of the latent node $X_1$ either because they directly influence $X_1$ (and hence $Y$) or because they act as intermediate nodes along a pathway originating from a variable upstream in the graph.
To this end, under each simulation scenario,
we fix the threshold for edge inclusion
$k^*=0.5$ and
include those edges $u\rightarrow 1$ such that $\hat{p}_{u\rightarrow 1}(\cdot) \ge 0.5$
in analogy with the median probability model of \citet{Barb:Berg:2004}.
The resulting 0-1 vector of indicators for edge inclusion is $\ba=(a_{1,1},\dots,a_{q,1})^\top$, where $a_{1,1} = 0$ while, for $u=2,\dots,q$, $a_{u,1}=1$ if $u\rightarrow 1$ is included, 0 otherwise.
Next we compute the proportion of predictors
that are correctly classified,
\be
p^*=\frac{1}{q-1}
\sum_{u=2}^{q}\mathbbm{1}\big\{a_{u,1}=\bA_{(u,1)}^{\D}\big\},
\ee
where $\bA_{(u,v)}^{\D}$ denotes the $(u,v)$
element of the adjacency matrix of $\D$.
The results are summarized in the box-plots
of Figure
\ref{fig:sim:boxplots:covariate:selection}
where we report the frequency distribution of $p^*$
computed over the 40 true DAGs.
While for $n=100$ the  proportion of
correctly classified edges presents some
variability with a median which is nevertheless around
80\% ($q=40$) and $90\%$ ($q=20$), the
performance greatly  improves as  the sample
size increases with  practically all
values being close to 1.

\vspace{0.3cm}

\begin{figure}
	\begin{center}
		\begin{tabular}{cc}
			$\quad\quad q = 20$ & $\quad\quad
			q = 40$ \\
			\vspace{0.1cm} \\
			
			\includegraphics[scale=0.22]{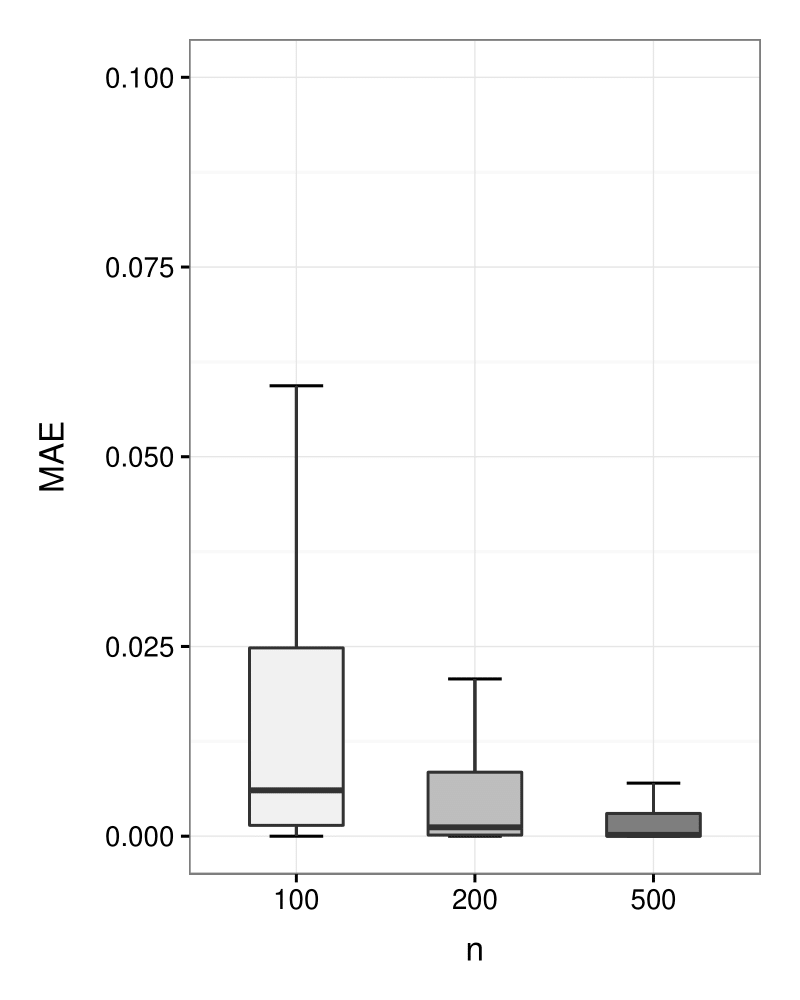}
			&
			
			\includegraphics[scale=0.22]{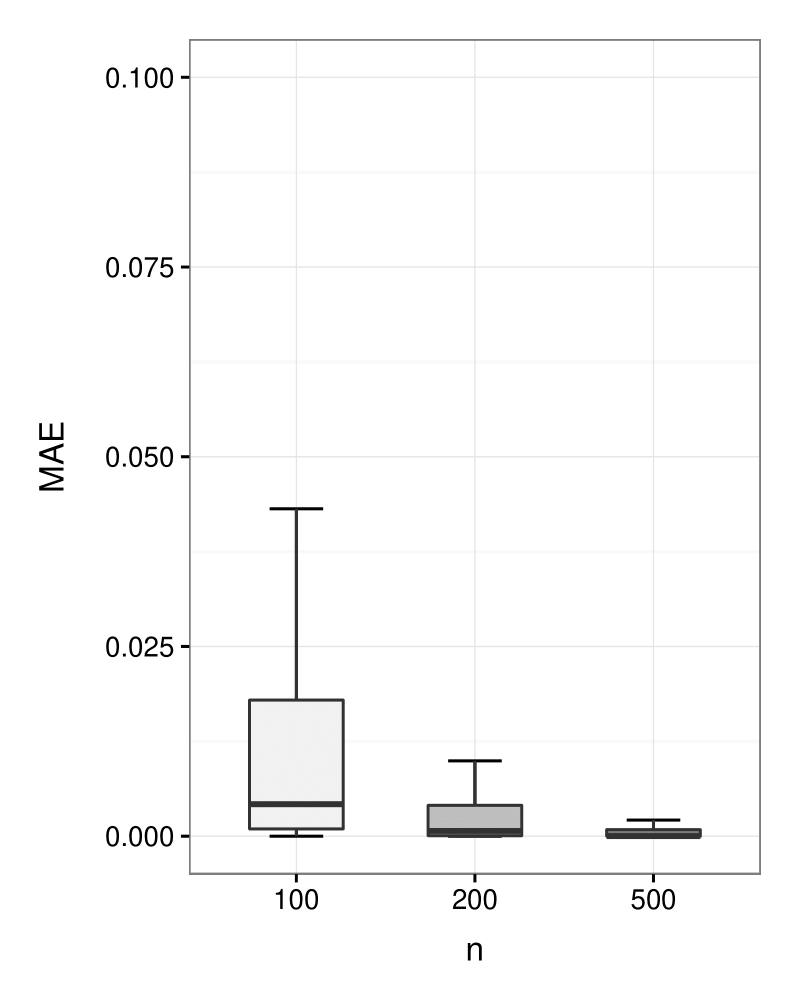}
		\end{tabular}
		\caption{\small Simulations.
			Distribution
			over $40$ datasets and
			nodes $s \in \{2,\dots,q\}$
			of the mean absolute error (MAE)
			of  BMA estimates of true causal effects.  Results are presented for each
			combination of number of nodes   $q\in\{20,40\}$ and sample size
			$n\in\{100,200,500\}$.}
		\label{fig:sim:boxplots}
	\end{center}
\end{figure}

We now focus on causal effect estimation.
Under each  simulated DAG $\D$ and parameters
$(\bD^{\D}, \bL^{\D})$ we first compute the
(true) covariance matrix $\bSigma^{\D}$ using
\eqref{eq:cholesky:fact}. Now recall from
\eqref{eq:causal:effect} that the causal
effect on $Y$  is a  probability  which also
depends on the level $\tilde x$ assigned to
the intervened variable $X_s$.
For each intervened node $s\in\{2,\dots,q\}$
we evaluate $\beta_s(\tilde{x},\bSigma^{\D},
\theta_0) \equiv \beta_s^{true}(\tilde{x})$
at each observed value of $X_s$ in the
simulation scenario, $(x_{1,s}, \dots,
x_{n,s})$, and obtain the $(n,1)$ vector of
causal effects
$\big(\beta_s^{true}(x_{1,s}),\dots,\beta_s^{true}(x_{n,s})\big)^\top$.
Next we produce the collection of BMA
estimates $\hat{\beta}_{s}^{BMA}(x_{1,s}),
\dots,$ $\hat{\beta}_{s}^{BMA}(x_{n,s})$
according to Equation
\eqref{eq:causal:effect:BMA}.
To evaluate the performance of our method in
estimating the  causal effect
we consider the differences
$\big(\beta_s^{true}(x_{i,s}) -
\hat{\beta}_{s}^{BMA}(x_{i,s})\big)$ and
compute the
mean absolute error (MAE)
\be
MAE_s =
\frac{1}{n}\sum_{i=1}^{n}\big|\beta_s^{true}(x_{i,s})
- \hat{\beta}_{s}^{BMA}(x_{i,s})\big|,
\ee
for each intervened node $s=2,\dots,q$.
Results are summarized in the box-plots of
Figure \ref{fig:sim:boxplots},  where we
report the distribution of the MAE
(constructed across the $40$ DAGs and nodes
$s=2,\dots,q$) as a function of $n$. As
expected, MAE decreases and approaches $0$ as
the sample size $n$ grows for both values of
$q$. Notice that the median value of MAE in
the worst case scenario $(q=20, n=100)$ is
about half of one percent.

\begin{figure}
	\begin{center}
		\begin{tabular}{c}
			\includegraphics[scale=0.22]{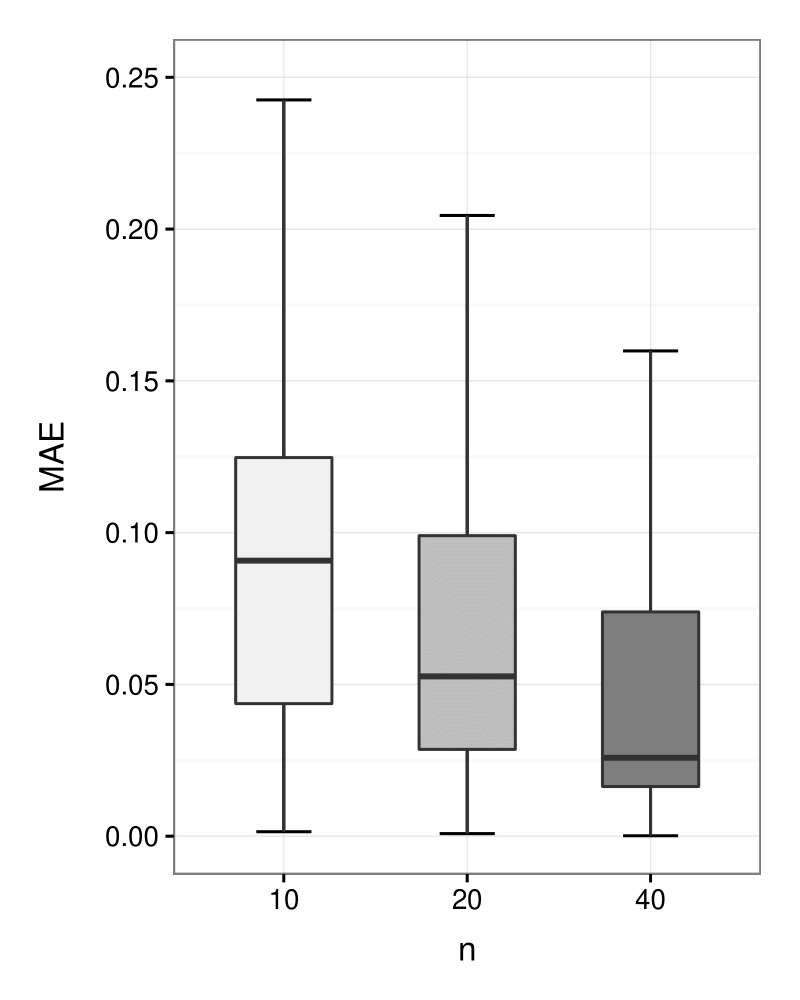}
		\end{tabular}
		\caption{\small Simulations. Distribution over 40 datasets and nodes $s\in\{2,\dots,q\}$ of the mean
			absolute error (MAE) of BMA estimates of true causal effects. Results are presented for number of nodes $q=40$ and sample size $n\in\{10,20,40\}$.}
		\label{fig:simulations:q>n}
	\end{center}
\end{figure}

Finally, we also explore settings where $n\le q$: in particular we include simulation results for $q=40$ and $n\in\{10,20,40\}$. Again, we generate $40$ DAGs and the allied parameters as in our first simulation study. Results are summarized in the box-plots of Figure \ref{fig:simulations:q>n}, where we report the distribution of the MAE, constructed across the 40 DAGs and nodes $s\in\{2,\dots,q\}$) as a function of $n$. It appears that, even if sample sizes are moderate, MAE decreases as $n$ grows.

\section{Analysis of gene expressions from breast cancer cells}
\label{sec:application}

In this section we apply our method to a gene expression dataset presented in \cite{Yin:et:al:2014}.
The aim of the original study was to evaluate the ability of a gene signature derived from breast cancer stem
cells to predict the risk of metastasis and survival in breast cancer patients.
To this end, a collection of genes which are believed to be the main responsible for tumor initiation,
progression, and response to therapy was considered. The study was motivated by  recent literature establishing the existence of a rare population of cells, called \textit{stem-like cells}, which supposedly  represent the cellular origin of cancer; see for instance \citet{O'Brien:et:al:2006}.
Gene-expression levels were measured on $n=198$ breast cancer patients of which $62$ manifested distant metastasis.
In the following we consider the expression levels of $q=28$ genes included in the original study and a binary response variable $Y$ indicating the occurrence (absence or presence, respectively $Y=0$ and $Y=1$) of distant metastasis. Evaluating the causal effect on $Y$ due to an hypothetical intervention on  a specific gene which sets its expression level may help understand which genes are particularly relevant for determining distant metastasis. This in turn can be useful to identify external  interventions, which are known to induce variations in the expression of specific genes.
For instance epigenetic modifications of gene expressions may be induced by lifestyle and environmental factors such as nutrition-dietary components, exercise, physical activity and toxins; see  \citet{Abdul:et:al:2017}. Similarly \citet{Campbell:et:al:2017}
examine the effects of lifestyle interventions on proposed biomarkers of lifestyle and cancer risk at the level of adipose tissue in humans.

\begin{figure}
	\begin{center}
		\includegraphics[scale=0.28]{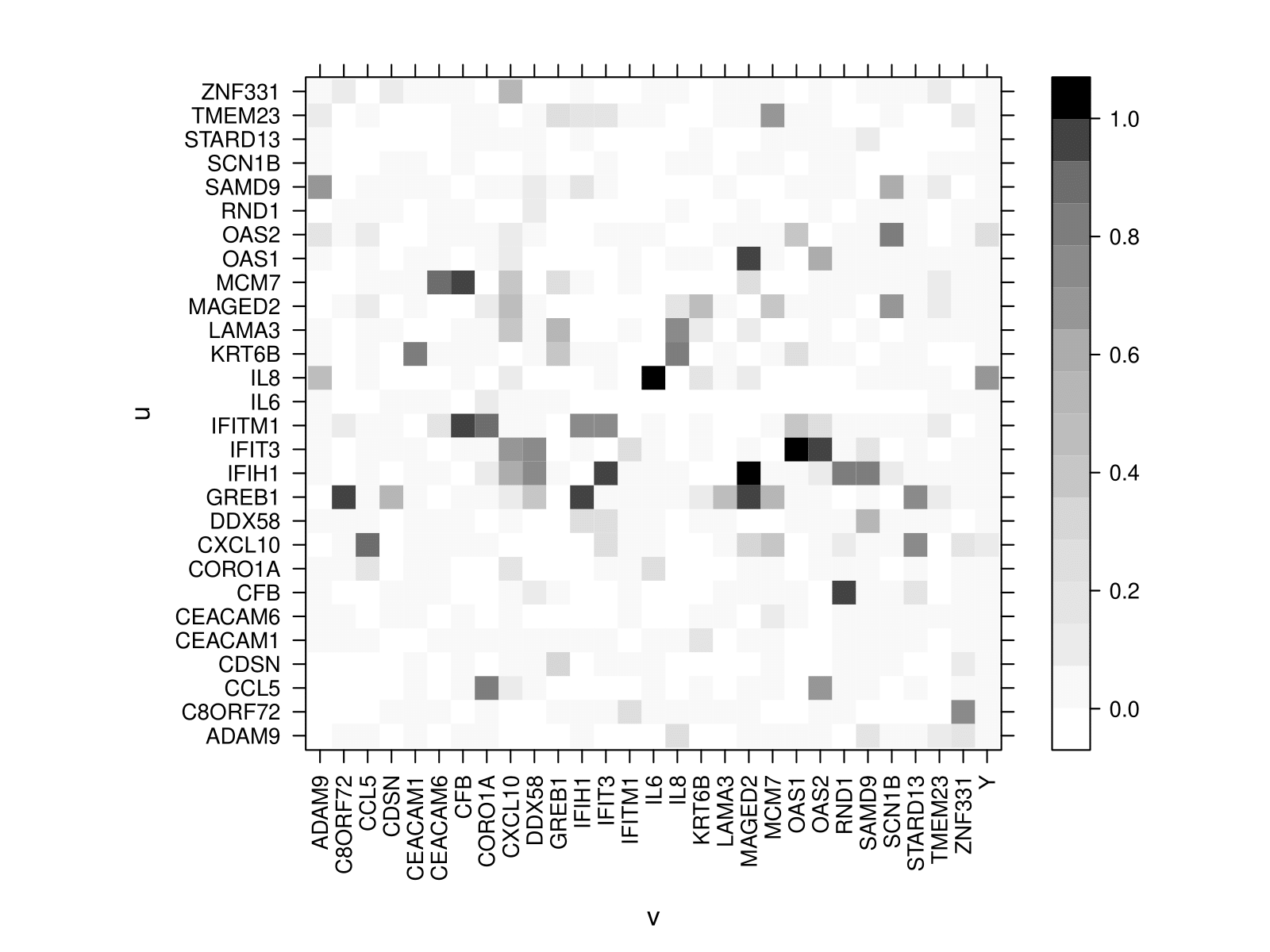}
		\caption{\small Gene expression data. Heat map with estimated marginal posterior probabilities of edge inclusion for each edge $u \rightarrow v$.}
		\label{fig:application:heatmap}
	\end{center}
\end{figure}

We apply Algorithm \ref{alg:MCMC} by fixing the number of MCMC iterations $T = 120000$. Observations from the continuous variables $X_2,\dots,X_q$ were standardized.
We also set $g=1/n$ and $a=q+1$ in the prior on the Cholesky parameters of $\bOmega$ as in the simulation scenarios of Section \ref{sec:simulations}.
We first use the MCMC output to estimate the posterior probability of inclusion of each directed edge $u\rightarrow v$, that we report in the heat map of Figure \ref{fig:application:heatmap}. Results show a substantial degree of sparsity in the underlying graph structure and only $48$ edges have a posterior probability of inclusion exceeding $0.5$.
Moreover, among the 28 genes, only gene IL8, for which $\hat{p}_{\textnormal{\,IL8}\rightarrow Y}(\cdot)=0.70$, seems to directly affect the response variable.

\begin{figure}
	\begin{center}
		\includegraphics[scale=0.28]{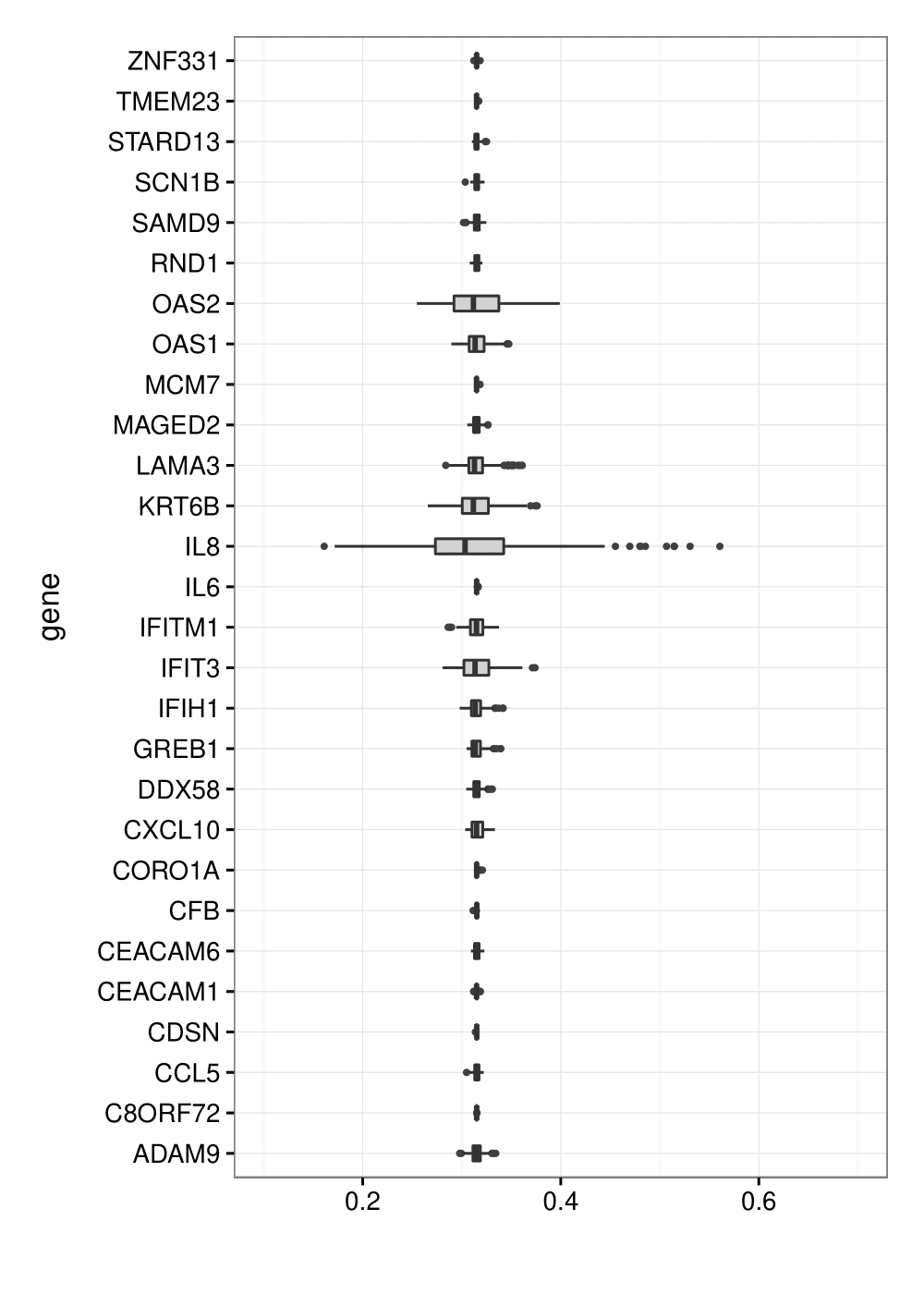}
		\caption{\small Gene expression data. Box-plots of  BMA estimate of causal effect. Each box-plot refers to one of the 28 genes $s$,  and represents the $n=198$ BMA estimates computed at each observed value  $(x_{1,s}, \dots, x_{n,s})$ of expression for gene $s$.}
		\label{fig:application:boxplot}
	\end{center}
\end{figure}

To evaluate the incidence of each gene on the probability of recurrence we  compute  the causal effect \eqref{eq:causal:effect}
on the response due to an intervention on a specific gene.
To this end, starting from the MCMC output we produce a BMA estimate $\hat\beta^{BMA}_s(\tilde x)$ for each gene $s=2,\dots,q$ according to \eqref{eq:causal:effect:BMA}. Since the causal effect depends on the level $\tilde x$ assigned to the intervened variable $X_s$, we evaluate $\hat{\beta}^{BMA}_s(\tilde{x})$ at each observed value of $X_s$, that is $x_{1,s}, \dots, x_{n,s}$. The results are reported in Figure \ref{fig:application:boxplot}, where each box-plot refers to a gene $s\in\{2,\dots,q\}$ and summarizes  the distribution of $n=198$ BMA estimates, $\big\{\hat\beta^{BMA}_s(x_{i,s})\big\}_{i=1}^{n}$. Because the data were standardized, the ranges of $X_2,\dots,X_q$ are similar and we can meaningfully compare  results across genes.

Recall from
Proposition \ref{proposition}
that $\gamma_s$
is the covariance between $X_s$ and $X_1$.
If $\gamma_s=0$,
Equation  \eqref{eq:causal:effect} shows  that
the causal effect on $Y$  due to an intervention on $X_s$ does not vary with $\tilde{x}$.
If prior information is  weak in relation to the  sample size,
the estimate of the causal effect will be close to the overall frequency  of distant metastasis in the sample (0.31).
This is the situation exhibited by most genes in Figure \ref{fig:application:boxplot}.
On the other hand  if $\gamma_s$ is not zero,  the collection of causal effects evaluated at
$x_{is}$, $i=1, \ldots,n$ will vary. Since  the observations are centred,
their average is zero and the causal effects
will be spread around the  value corresponding  to the average $\bar{x}_s=0$ whose estimate  is 0.31 as indicated above.
This is what happens for a few genes such as  IL8, OAS2 and KRT6B, which exhibit a much greater  variability of the causal effect across their measurements, implying that their regulation can influence  the  occurrence of distant metastasis.
In particular, gene IL8 has also  been identified as having a potential impact  on  cancer cells in several studies
\citep{Waugh:et:al:2008}.
Other  genes which stand out in terms of variability are  OAS2 and KRT6B, with the latter not directly linked to $Y$ (as one can see from the heat map of Figure \ref{fig:application:heatmap})
and exhibiting a moderate causal effect on $Y$
which is likely due to the strong association of KRT6B with IL8 (as it emerges from the posterior probabilities of edge inclusion in Figure \ref{fig:application:heatmap}).

\begin{figure}
	\begin{center}
		\includegraphics[scale=0.22]{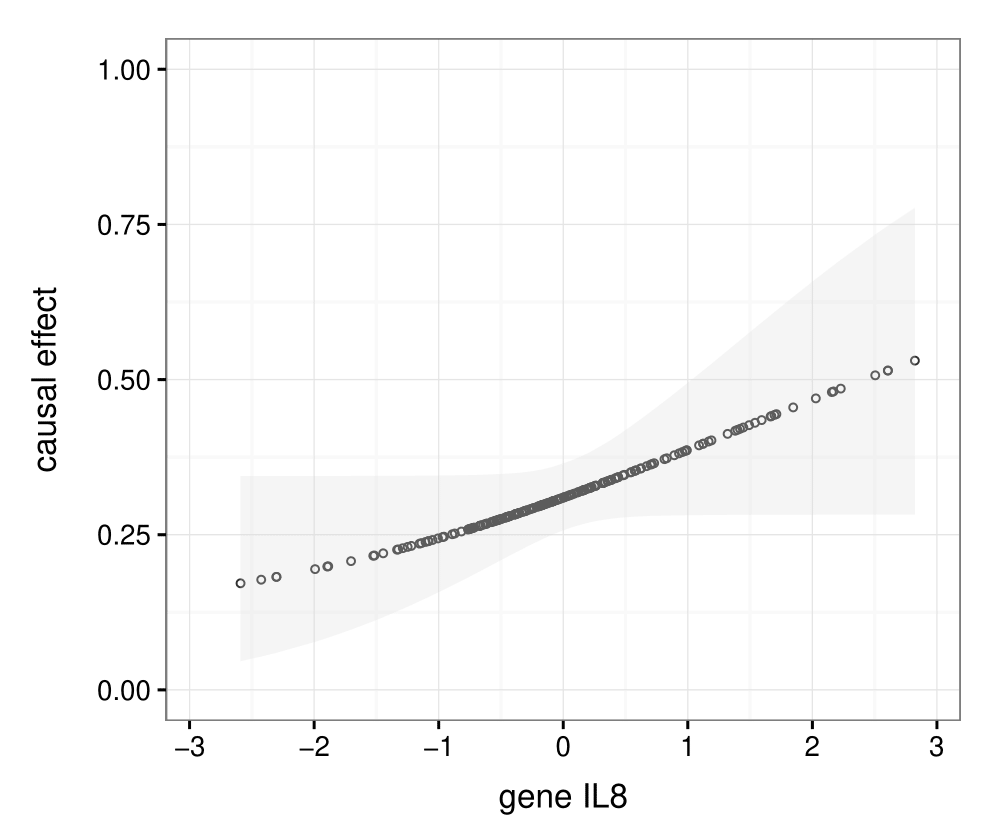}
		\includegraphics[scale=0.22]{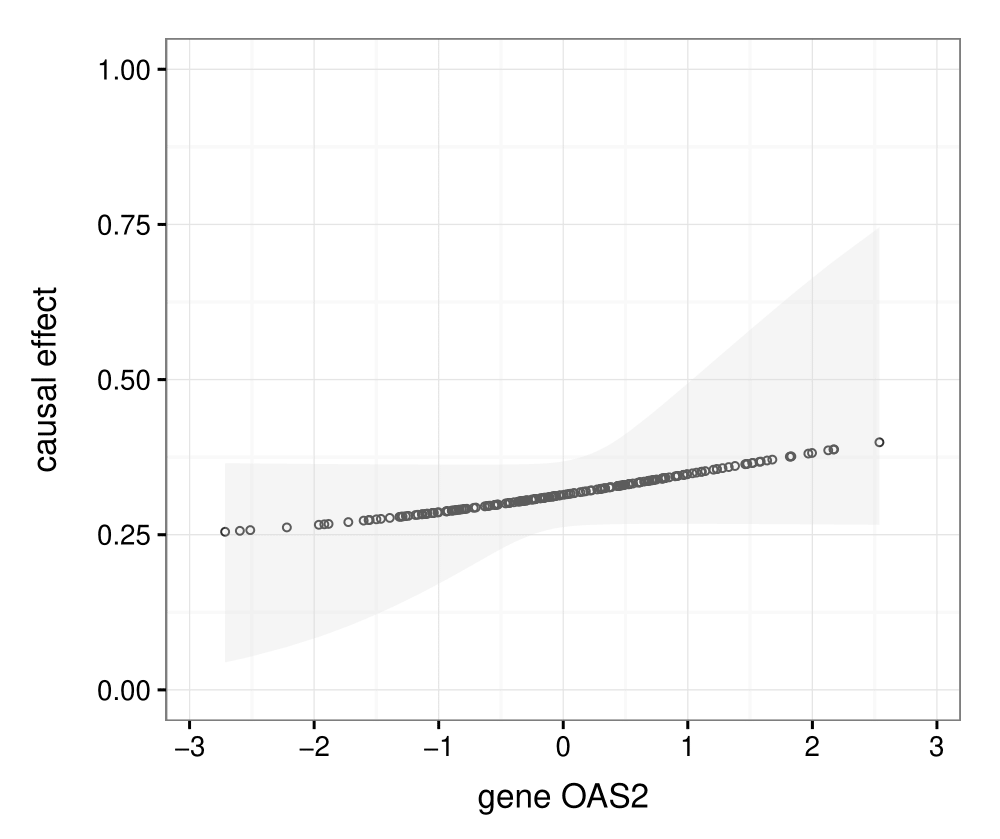}
		\caption{\small Gene expression data. BMA estimates (dots) and credible regions at level $95\%$ (grey area) for two selected genes, IL8 and OAS2.}
		\label{fig:application:causals}
	\end{center}
\end{figure}

For genes IL8 and OAS2 we also report in Figure \ref{fig:application:causals} more detailed results for causal effect estimation. In particular, each plot reports the BMA estimates $\big\{\hat\beta^{BMA}_s(x_{i,s})\big\}_{i=1}^{n}$ 
(represented by $n=198$ dots),  and the corresponding credible regions at level $95\%$ represented by the grey area. Results show that increasing expression levels of IL8 are likely to increase the presence of distant metastasis, with BMA estimates of the probability of recurrence ranging in the interval $[0.18; 0.54]$. This is consistent with results we have obtained showing that most of the mass of the distribution of the coefficient $\gamma_s$ for these genes is assigned to the positive half-line; see also the discussion after \eqref{eq:causal:effect}. Moreover, more extreme levels of IL8 are associated with larger credible regions. A similar behavior, although less pronounced, is observed for gene OAS2 with BMA estimates ranging between $[0.25; 0.41]$.

\section{Discussion}
\label{sec:discussion}

In this paper we  deal with causal effect estimation from observational data.
We consider a system of real-valued variables together with a binary response;  interest lies in the  evaluation of the causal effect on the response due to an external intervention on  a variable.
We assume that the response is generated by standard thresholding applied to a continuous latent variable, and assume that the joint distribution of all continuous variables belongs to  a zero-mean Gaussian Directed Acyclic Graph (DAG) model, or equivalently a Structural Equation Model (SEM),  and name the resulting model \emph{DAG-probit}.

Rather than constructing a prior distribution on the covariance (or precision) matrix constrained by a given DAG, we proceed by assigning to the corresponding  Cholesky parameters a DAG-Wishart prior whose hyperparameters are deduced from a single unique Wishart distribution. This technique not only drastically simplifies prior elicitation, but has the important advantage of producing score equivalence, meaning that  marginal likelihoods of Markov equivalent DAG models are all equal \citep{Geig:Heck:2002}.
This feature will have a useful implication, as we discuss at the end of this section.


Because the structure of the data-generating DAG is unknown, we construct an MCMC sampler
whose target is the joint posterior distribution of the DAG and the allied Cholesky parameters.
This is achieved by carefully tailoring
a Partial Analytic Structure (PAS) algorithm to our DAG setting.
As a by-product,  we recover the MCMC sequence of causal effects corresponding to  each visited DAG; this represents the input to our final Bayesian Model Averaging (BMA) estimate, which naturally accounts for model uncertainty on the underlying graph structure.

The assumption of jointly normally  distributed random variables can be a source of concern whenever one faces a concrete data analysis. With regard to our application the Gaussian assumption has been often used to analyse gene expression data; see for instance \citet{Dobra:etal:2004} and \citet{Markowetx:Spang:2007}.
In addition, 
it allows to easily incorporate the binary outcome through a latent component, and  results in an efficient algorithm, because of closed-form expressions  both for the   posterior distribution of parameters,  as well as for the marginal likelihood of models.

Besides the assumption of normality, our model posits a unique graphical structure as the generating mechanism of all observations.
Nevertheless, some problems may suggest to partition the units into groups each having
a specific graphical structure which can be however related to the other ones, as in  gene expressions collected on multiple tissues from the same individual \citep{Xie:etal:2017}.
In this setting a  multiple graphical model setup would be more appropriate to encourage  similarities between group graphical structures; see for instance \citet{Pete:etal:2015} for a Bayesian analysis of multiple  Gaussian undirected graphical models.
The latter  could be a useful starting point for an extension of our DAG-probit model to multiple groups.
%

In this work we consider causal effects as obtained from interventions on single nodes.
However in practice an exogenous intervention may affect many variables (genes) simultaneously
and accordingly one may want to predict for instance the effect of a double or triple gene knockout on the response.
Causal effect estimation from joint interventions is carried out in a Gaussian setting by \citet{Nandy:Maathuis:Richardson:2017} using a frequentist approach.
Their results show that the causal effect of $X_s$ on the response in a joint intervention on a given set of variables can be still expressed as a function of the covariance matrix Markov w.r.t. $\D$.
The same problem can be tackled by adopting a Bayesian methodology which combines DAG structural learning and causal effect estimation and is currently under investigation by ourselves. In addition, an extension to DAG-probit models should be feasible along the lines of this paper.

The methodology adopted in this work
revolves around
DAGs. However, it is known that in the Gaussian setting  DAGs encoding the same conditional independencies (Markov equivalent DAGs) are not distinguishable using observational data \citep{Verm:Pear:1990}
and can be collected into Markov equivalence classes (MECs). Accordingly, when the goal of the analysis  is  structural learning (model selection) MECs represent the  appropriate inferential object \citep{Ande:etal:1997}.
However, if  the objective is causal effect estimation, this is no longer so, because Markov equivalent DAGs may return distinct causal effects. An inspection of \eqref{eq:causal:effect} reveals the reason:  a causal effect depends on the parent set of the intervened node, and  this may differ  among DAGs within the same MEC.
Yet MECs  can  be exploited also for causal inference, as we now detail.
In a frequentist setting,  \citet{Maat:etal:2009}  first estimate a MEC using the classic PC algorithm \citep{Spir:Glym:Sche:2000}, and then provide an estimate of the causal effect under each DAG within the estimated equivalence class.
Alternatively, a Bayesian methodology would first determine the posterior distribution on MEC space,  and then, conditionally on a given MEC, compute the posterior of each causal effect within the class (one for each DAG). A single MEC causal effect  estimate can  be obtained by averaging effects across DAGs,  using   uniform weights on equivalent DAGs.  Finally,  an overall Bayesian Model Averaging (BMA) estimate  can be obtained by averaging MEC-conditional estimates using posterior probabilities of  MECs as weights; for details see  \citep{Caste:Cons:2020}.
We remark that the  above strategies require an exhaustive enumeration of all DAGs belonging to a MEC, which is not feasible even for a moderate number of nodes. Accordingly one considers only  the distinct causal effects within a given MEC,  because these  values can be efficiently recovered \citep[Algorithm 3]{Maat:etal:2009} even in high-dimensional settings.
In this work we seemingly ignore the issue of DAG Markov equivalence,  and propose a causal inference procedure  which directly focuses on DAG space, rather than MEC space.
However, as already remarked at the beginning of this section,  our method  for parameter prior construction across DAG models guarantees score equivalence for DAGs within the same MEC. This, together with a uniform prior on DAGs within the same MEC,  ensures that causal effects associated to Markov equivalent DAGs will be assigned equal weights  in the resulting BMA estimate.
\black


\bibliographystyle{biometrika}
\bibliography{biblio}

\begin{thebibliography}{33}
\expandafter\ifx\csname natexlab\endcsname\relax\def\natexlab#1{#1}\fi
\expandafter\ifx\csname url\endcsname\relax
  \def\url#1{\texttt{#1}}\fi
\expandafter\ifx\csname urlprefix\endcsname\relax\def\urlprefix{URL }\fi
\providecommand{\eprint}[2][]{\url{#2}}

\bibitem[{Abdul et~al.(2017)Abdul, Yu, Chung, Jung \& J.S.}]{Abdul:et:al:2017}
\textsc{Abdul, Q.}, \textsc{Yu, B.}, \textsc{Chung, H.}, \textsc{Jung, H.} \&
  \textsc{J.S., C.} (2017).
\newblock Epigenetic modifications of gene expression by lifestyle and
  environment.
\newblock \textit{Archives of Pharmacal Research} 40 1219--1237.
\newblock \urlprefix\url{https://doi.org/10.1007/s12272-017-0973-3}.

\bibitem[{Albert \& Chib(1993)}]{Albert:Chib:1993}
\textsc{Albert, J.~H.} \& \textsc{Chib, S.} (1993).
\newblock Bayesian analysis of binary and polychotomous response data.
\newblock \textit{Journal of the American Statistical Association} 88 669--679.
\newblock \urlprefix\url{http://www.jstor.org/stable/2290350}.

\bibitem[{Andersson et~al.(1997)Andersson, Madigan \& Perlman}]{Ande:etal:1997}
\textsc{Andersson, S.~A.}, \textsc{Madigan, D.} \& \textsc{Perlman, M.~D.}
  (1997).
\newblock A characterization of {M}arkov equivalence classes for acyclic
  digraphs.
\newblock \textit{The Annals of Statistics} 25 505--541.
\newblock \urlprefix\url{http://dx.doi.org/10.1214/aos/1031833662}.

\bibitem[{Barbieri \& Berger(2004)}]{Barb:Berg:2004}
\textsc{Barbieri, M.~M.} \& \textsc{Berger, J.~O.} (2004).
\newblock Optimal predictive model selection.
\newblock \textit{The Annals of Statistics} 32 870--897.
\newblock \urlprefix\url{https://doi.org/10.1214/009053604000000238}.

\bibitem[{Ben-David et~al.(2015)Ben-David, Li, Massam \&
  Rajaratnam}]{Ben:Massam:arXiv}
\textsc{Ben-David, E.}, \textsc{Li, T.}, \textsc{Massam, H.} \&
  \textsc{Rajaratnam, B.} (2015).
\newblock High dimensional {B}ayesian inference for {G}aussian directed acyclic
  graph models.
\newblock \textit{arXiv pre-print}
  \urlprefix\url{https://arxiv.org/abs/1109.4371}.

\bibitem[{Campbell et~al.(2017)Campbell, Landells, Fan \&
  Brenner}]{Campbell:et:al:2017}
\textsc{Campbell, K.~L.}, \textsc{Landells, C.~E.}, \textsc{Fan, J.} \&
  \textsc{Brenner, D.~R.} (2017).
\newblock A systematic review of the effect of lifestyle interventions on
  adipose tissue gene expression: Implications for carcinogenesis.
\newblock \textit{Obesity} 25 S40--S51.
\newblock
  \urlprefix\url{https://onlinelibrary.wiley.com/doi/abs/10.1002/oby.22010}.

\bibitem[{Cao et~al.(2019)Cao, Khare \& Ghosh}]{cao:et:al:2019}
\textsc{Cao, X.}, \textsc{Khare, K.} \& \textsc{Ghosh, M.} (2019).
\newblock Posterior graph selection and estimation consistency for
  high-dimensional {B}ayesian {DAG} models.
\newblock \textit{The Annals of Statistics} 47 319--348.
\newblock \urlprefix\url{https://doi.org/10.1214/18-AOS1689}.

\bibitem[{Castelletti \& Consonni(2020{\natexlab{a}})}]{Caste:Cons:2020}
\textsc{Castelletti, F.} \& \textsc{Consonni, G.} (2020{\natexlab{a}}).
\newblock {B}ayesian inference of causal effects from observational data in
  {G}aussian graphical models.
\newblock \textit{Biometrics, In press} .

\bibitem[{Castelletti \& Consonni(2020{\natexlab{b}})}]{caste:cons:suppl:BA}
\textsc{Castelletti, F.} \& \textsc{Consonni, G.} (2020{\natexlab{b}}).
\newblock Supplementary material to ``{B}ayesian causal inference in probit
  graphical models" .

\bibitem[{Dobra et~al.(2004)Dobra, Hans, Jones, Nevins, Yao \&
  West}]{Dobra:etal:2004}
\textsc{Dobra, A.}, \textsc{Hans, C.}, \textsc{Jones, B.}, \textsc{Nevins,
  J.~R.}, \textsc{Yao, G.} \& \textsc{West, M.} (2004).
\newblock Sparse graphical models for exploring gene expression data.
\newblock \textit{Journal of Multivariate Analysis} 90 196 -- 212.
\newblock
  \urlprefix\url{http://www.sciencedirect.com/science/article/pii/S0047259X04000259}.

\bibitem[{Friedman(2004)}]{Frie:2004}
\textsc{Friedman, N.} (2004).
\newblock Inferring cellular networks using probabilistic graphical models.
\newblock \textit{Science} 303 799--805.
\newblock \urlprefix\url{https://science.sciencemag.org/content/303/5659/799}.

\bibitem[{Friedman \& Koller(2003)}]{Frie:Koll:2003}
\textsc{Friedman, N.} \& \textsc{Koller, D.} (2003).
\newblock Being {B}ayesian about network structure. {A} {B}ayesian approach to
  structure discovery in {B}ayesian networks.
\newblock \textit{Machine Learning} 50 95--125.
\newblock \urlprefix\url{https://doi.org/10.1023/A:1020249912095}.

\bibitem[{Garc\'{i}a-Donato \&
  Mart\'{i}nez-Beneito(2013)}]{Garcia:Donato:et:al:2013}
\textsc{Garc\'{i}a-Donato, G.} \& \textsc{Mart\'{i}nez-Beneito, M.~A.} (2013).
\newblock On sampling strategies in bayesian variable selection problems with
  large model spaces.
\newblock \textit{Journal of the American Statistical Association} 108
  340--352.
\newblock \urlprefix\url{https://doi.org/10.1080/01621459.2012.742443}.

\bibitem[{Geiger \& Heckerman(2002)}]{Geig:Heck:2002}
\textsc{Geiger, D.} \& \textsc{Heckerman, D.} (2002).
\newblock Parameter priors for directed acyclic graphical models and the
  characterization of several probability distributions.
\newblock \textit{The Annals of Statistics} 30 1412--1440.
\newblock \urlprefix\url{https://doi.org/10.1214/aos/1035844981}.

\bibitem[{Godsill(2012)}]{Godsill:2012}
\textsc{Godsill, S.~J.} (2012).
\newblock On the relationship between markov chain monte carlo methods for
  model uncertainty.
\newblock \textit{Journal of Computational and Graphical Statistics} 10
  230--248.
\newblock \urlprefix\url{https://doi.org/10.1198/10618600152627924}.

\bibitem[{Guo et~al.(2015)Guo, Levina, Michailidis \&
  Zhu}]{Guo:Levina:et:al:2015}
\textsc{Guo, J.}, \textsc{Levina, E.}, \textsc{Michailidis, G.} \& \textsc{Zhu,
  J.} (2015).
\newblock Graphical models for ordinal data.
\newblock \textit{Journal of Computational and Graphical Statistics} 24
  183--204.
\newblock \urlprefix\url{https://doi.org/10.1080/10618600.2014.889023}.

\bibitem[{Lauritzen(1996)}]{Laur:1996}
\textsc{Lauritzen, S.~L.} (1996).
\newblock \textit{Graphical Models}.
\newblock Oxford University Press.

\bibitem[{Maathuis \& Nandy(2016)}]{Maathuis:Nandy:Review}
\textsc{Maathuis, M.} \& \textsc{Nandy, P.} (2016).
\newblock A review of some recent advances in causal inference.
\newblock In P.~B{\"u}hlmann, P.~Drineas, M.~Kane \& M.~van~der Laan, eds.,
  \textit{Handbook of Big Data}. Chapman and Hall/CRC, 387--408.

\bibitem[{Maathuis et~al.(2009)Maathuis, Kalisch \& Bühlmann}]{Maat:etal:2009}
\textsc{Maathuis, M.~H.}, \textsc{Kalisch, M.} \& \textsc{Bühlmann, P.}
  (2009).
\newblock Estimating high-dimensional intervention effects from observational
  data.
\newblock \textit{The Annals of Statistics} 37 3133--3164.
\newblock \urlprefix\url{https://doi.org/10.1214/09-AOS685}.

\bibitem[{Markowetz \& Spang(2007)}]{Markowetx:Spang:2007}
\textsc{Markowetz, F.} \& \textsc{Spang, R.} (2007).
\newblock Inferring cellular networks - a review.
\newblock \textit{BMC Bioinformatics} 8.
\newblock \urlprefix\url{https://doi.org/10.1186/1471-2105-8-S6-S5}.

\bibitem[{Nandy et~al.(2017)Nandy, Maathuis \&
  Richardson}]{Nandy:Maathuis:Richardson:2017}
\textsc{Nandy, P.}, \textsc{Maathuis, M.~H.} \& \textsc{Richardson, T.~S.}
  (2017).
\newblock Estimating the effect of joint interventions from observational data
  in sparse high-dimensional settings.
\newblock \textit{Ann. Statist.} 45 647--674.
\newblock \urlprefix\url{https://doi.org/10.1214/16-AOS1462}.

\bibitem[{O’Brien et~al.(2006)O’Brien, Pollett, Gallinger \&
  Dick}]{O'Brien:et:al:2006}
\textsc{O’Brien, C.~A.}, \textsc{Pollett, A.}, \textsc{Gallinger, S.} \&
  \textsc{Dick, J.~E.} (2006).
\newblock A human colon cancer cell capable of initiating tumour growth in
  immunodeficient mice.
\newblock \textit{Nature} 445 106--110.
\newblock \urlprefix\url{https://doi.org/10.1038/nature05372}.

\bibitem[{Pearl(1995)}]{Pearl:1995}
\textsc{Pearl, J.} (1995).
\newblock Causal diagrams for empirical research.
\newblock \textit{Biometrika} 82 669--688.
\newblock \urlprefix\url{http://www.jstor.org/stable/2337329}.

\bibitem[{Pearl(2000)}]{Pear:2000}
\textsc{Pearl, J.} (2000).
\newblock \textit{Causality: Models, Reasoning, and Inference}.
\newblock Cambridge University Press, Cambridge.

\bibitem[{Pearl(2009)}]{Pearl:2009:Surveys}
\textsc{Pearl, J.} (2009).
\newblock Causal inference in statistics: An overview.
\newblock \textit{Statistics Surveys} 3 96--146.
\newblock \urlprefix\url{https://doi.org/10.1214/09-SS057}.

\bibitem[{Peters \& B{\"u}hlmann(2014)}]{Pete:Buhl:2014}
\textsc{Peters, J.} \& \textsc{B{\"u}hlmann, P.} (2014).
\newblock Identifiability of {G}aussian structural equation models with equal
  error variances.
\newblock \textit{Biometrika} 101 219--228.
\newblock \urlprefix\url{http://www.jstor.org/stable/43305605}.

\bibitem[{Peterson et~al.(2015)Peterson, Stingo \& Vannucci}]{Pete:etal:2015}
\textsc{Peterson, C.}, \textsc{Stingo, F.~C.} \& \textsc{Vannucci, M.} (2015).
\newblock Bayesian inference of multiple {G}aussian graphical models.
\newblock \textit{Journal of the American Statistical Association} 110
  159--174.
\newblock PMID: 26078481,
  \urlprefix\url{https://doi.org/10.1080/01621459.2014.896806}.

\bibitem[{Spirtes et~al.(2000)Spirtes, Glymour \&
  Scheines}]{Spir:Glym:Sche:2000}
\textsc{Spirtes, P.}, \textsc{Glymour, C.} \& \textsc{Scheines, R.} (2000).
\newblock \textit{Causation, Prediction and Search (2nd edition)}.
\newblock Cambridge, MA: The MIT Press.

\bibitem[{Verma \& Pearl(1990)}]{Verm:Pear:1990}
\textsc{Verma, T.} \& \textsc{Pearl, J.} (1990).
\newblock Equivalence and synthesis of causal models.
\newblock In \textit{Proceedings of the Sixth Annual Conference on Uncertainty
  in Artificial Intelligence}, UAI 90. New York, NY, USA: Elsevier Science
  Inc., 255--270.

\bibitem[{Wang \& Li(2012)}]{wang:li:2012}
\textsc{Wang, H.} \& \textsc{Li, S.~Z.} (2012).
\newblock Efficient gaussian graphical model determination under g -wishart
  prior distributions.
\newblock \textit{Electronic Journal of Statistics} 6 168--198.
\newblock \urlprefix\url{https://doi.org/10.1214/12-EJS669}.

\bibitem[{Waugh \& Wilson(2008)}]{Waugh:et:al:2008}
\textsc{Waugh, D.~J.} \& \textsc{Wilson, C.} (2008).
\newblock The interleukin-8 pathway in cancer.
\newblock \textit{Clinical Cancer Research} 14 6735--6741.
\newblock \urlprefix\url{https://doi.org/10.1158/1078-0432.CCR-07-4843}.

\bibitem[{Xie et~al.(2017)Xie, Liu \& Valdar}]{Xie:etal:2017}
\textsc{Xie, Y.}, \textsc{Liu, Y.} \& \textsc{Valdar, W.} (2017).
\newblock {Joint estimation of multiple dependent Gaussian graphical models
  with applications to mouse genomics}.
\newblock \textit{Biometrika} 103 493--511.
\newblock \urlprefix\url{https://doi.org/10.1093/biomet/asw035}.

\bibitem[{Yin et~al.(2014)Yin, Liu, Xu, Yu, Ding, Yang, Tang, Liu, Ma, Xia, Lin
  \& Wang}]{Yin:et:al:2014}
\textsc{Yin, Z.-Q.}, \textsc{Liu, J.-J.}, \textsc{Xu, Y.-C.}, \textsc{Yu, J.},
  \textsc{Ding, G.-H.}, \textsc{Yang, F.}, \textsc{Tang, L.}, \textsc{Liu,
  B.-H.}, \textsc{Ma, Y.}, \textsc{Xia, Y.-W.}, \textsc{Lin, X.-L.} \&
  \textsc{Wang, H.-X.} (2014).
\newblock A 41-gene signature derived from breast cancer stem cells as a
  predictor of survival.
\newblock \textit{Journal of Experimental \& Clinical Cancer Research} 33.
\newblock \urlprefix\url{https://doi.org/10.1186/1756-9966-33-49}.

\end{thebibliography}


\begin{thebibliography}{3}
\expandafter\ifx\csname natexlab\endcsname\relax\def\natexlab#1{#1}\fi
\expandafter\ifx\csname url\endcsname\relax
  \def\url#1{\texttt{#1}}\fi
\expandafter\ifx\csname urlprefix\endcsname\relax\def\urlprefix{URL }\fi
\providecommand{\eprint}[2][]{\url{#2}}

\bibitem[{Godsill(2012)}]{Godsill:2012}
\textsc{Godsill, S.~J.} (2012).
\newblock On the relationship between markov chain monte carlo methods for
  model uncertainty.
\newblock \textit{Journal of Computational and Graphical Statistics} 10
  230--248.
\newblock \urlprefix\url{https://doi.org/10.1198/10618600152627924}.

\bibitem[{Melchers \& Beck(2017)}]{Beck:Melchers:2017}
\textsc{Melchers, R.~E.} \& \textsc{Beck, A.~T.} (2017).
\newblock \textit{Structural Reliability Analysis and Prediction}.
\newblock John Wiley \& Sons, Ltd.

\bibitem[{Wang \& Li(2012)}]{wang:li:2012}
\textsc{Wang, H.} \& \textsc{Li, S.~Z.} (2012).
\newblock Efficient gaussian graphical model determination under g -wishart
  prior distributions.
\newblock \textit{Electronic Journal of Statistics} 6 168--198.
\newblock \urlprefix\url{https://doi.org/10.1214/12-EJS669}.

\end{thebibliography}

\end{document}


\maketitle

\section{Proof of Proposition 3.1}

\label{subsec:Appendix:causal}
In this section we give a proof of Proposition 3.1. To this end we first introduce the following two lemmata.

\begin{lemma}
	\label{corollary_1}
	Let $\bx,\bb \in \Re^d$ be two vectors, $\bM\in\Re^{d\times d}$ a symmetric and invertible matrix. Then
	\be
	\bx^\top\bM\bx - 2\bb^\top\bx=(\bx-\bM^{-1}\bb)^\top\bM(\bx-\bM^{-1}\bb)-\bb^\top\bM^{-1}\bb.
	\ee
\end{lemma}

\begin{proof}
	Simply expand the right-hand side of the equation.
\end{proof}
\vspace{0.2cm}

\begin{lemma}
	\label{corollary_2}
	Let $V\g (\boldsymbol{U}=\bu) \sim \N(\mu+\bgamma^\top\bu,\delta^2)$, $\boldsymbol{U}\sim\N_d(\bzero,\bSigma)$. Then
	\be
	V &\sim& \N\left(\mu, \frac{\delta^2}{1-(\bgamma^\top\bT^{-1}\bgamma)/\delta^2}\right),
	\ee
	where $\bT=\bSigma^{-1}+\bgamma\bgamma^\top/\delta^2$.
	
	\begin{proof}
		We can write by definition
		\be
		f(v)&=&\int f(v\g\bu)f(\bu) \, d\bu \nonumber \\
		&=& \int_{\Re^d}\frac{1}{\sqrt{2\pi}\delta}
		\exp\left\{-\frac{1}{2\delta^2}(v-\mu-\bgamma^\top\bu)^2\right\} \\
		&\cdot&
		\frac{1}{(2\pi)^{d/2}|\bSigma|^{1/2}}\exp\left\{-\frac{1}{2}\bu^\top\bSigma^{-1}\bu\right\} \, d\bu.
		\ee
		Next, observe that
		\be
		-\frac{1}{2\delta^2}(v-\mu-\bgamma^\top\bu)^2 &=& -\frac{1}{2\delta^2}\left[(v-\mu)^2+(\bgamma^\top\bu)^2-2(v-\mu)\bgamma^\top\bu\right] \\
		&=&
		-\frac{1}{2\delta^2}\left[(v-\mu)^2+\bu^\top\bgamma\bgamma^\top\bu-2(v-\mu)\bgamma^\top\bu\right] \\
		&=&
		-\frac{1}{2\delta^2}\left[(v-\mu)^2+\bu^\top\bGamma\,\bu-2(v-\mu)\bgamma^\top\bu\right],
		\ee
		being $\bGamma=\bgamma\bgamma^\top$.
		Therefore we can write
		\be
		f(v) &=& \frac{1}{\sqrt{2\pi}\delta} \exp\left\{-\frac{1}{2}\left(\frac{v-\mu}{\delta}\right)^2\right\} \\
		&\cdot&
		\frac{1}{|\bSigma|^{1/2}}
		\int_{\Re^{d}}\frac{1}{(2\pi)^{d/2}}
		\exp\left\{-\frac{1}{2}\left[\bu^\top\left(\bSigma^{-1}+\frac{\bGamma}{\delta^2}\right)\bu
		-2\frac{(v-\mu)}{\delta^2}\bgamma^\top\bu\right]\right\} \, d\bu.
		\ee
		
		\vspace{0.3cm}
		
		\noindent
		Let now $\bT=\bSigma^{-1}+\bGamma/\delta^2$. By Lemma \ref{corollary_1} we obtain
		\be
		\bu^\top\bT\,\bu-2\left[\frac{(v-\mu)}{\delta^2}\bgamma^\top\right]\bu
		&=& \left[\bu-\bT^{-1}\frac{(v-\mu)}{\delta^2}\bgamma\right]^\top\bT
		\left[\bu-\bT^{-1}\frac{(v-\mu)}{\delta^2}\bgamma\right] \\
		&-& \left[\frac{(v-\mu)}{\delta^2}\bgamma^\top\right]\bT^{-1}\left[\frac{(v-\mu)}{\delta^2}\bgamma\right].
		\ee
		Therefore,
		\be
		f(v) &=& \frac{1}{\sqrt{2\pi}\delta} \exp\left\{-\frac{1}{2}\left(\frac{v-\mu}{\delta}\right)^2\right\} \\
		&\cdot&
		\frac{1}{|\bSigma|^{1/2}}
		\int_{\Re^{d}}\frac{1}{(2\pi)^{d/2}}
		\exp\left\{-\frac{1}{2}\left[\left(\bu-\bT^{-1}\frac{(v-\mu)}{\delta^2}\bgamma\right)^\top\bT
		\left(\bu-\bT^{-1}\frac{(v-\mu)}{\delta^2}\bgamma\right)\right]\right\} \\
		&\cdot&
		\exp \left\{\,\frac{1}{2}\left[\frac{(v-\mu)}{\delta^2}\bgamma^\top\right]\bT^{-1}\left[\frac{(v-\mu)}{\delta^2}\bgamma\right]\right\}
		\, d\bu.
		\ee
		Moreover we can write
		\be
		f(v) &=&
		\frac{1}{\sqrt{2\pi}\delta} \exp\left\{-\frac{1}{2}\left(\frac{v-\mu}{\delta}\right)^2\right\} \cdot
		\frac{1}{|\bSigma|^{1/2}}
		\exp \left\{\,\frac{1}{2}\left[\frac{(v-\mu)}{\delta^2}\bgamma^\top\right]\bT^{-1}\left[\frac{(v-\mu)}{\delta^2}\bgamma\right]\right\} \\
		&\cdot&
		\frac{1}{|\bT|^{1/2}}
		\int_{\Re^d}\frac{|\bT|^{1/2}}{(2\pi)^{d/2}}
		\exp\left\{-\frac{1}{2}\left(\bu-\frac{(v-\mu)}{\delta^2}\,\bT^{-1}\bgamma\right)^\top
		\bT\left(\bu-\frac{(v-\mu)}{\delta^2}\,\bT^{-1}\bgamma\right)\right\} \, d\bu.
		\ee
		Hence,
		\be
		f(v) &=&
		\frac{1}{\sqrt{2\pi}\delta}
		\frac{1}{|\bSigma|^{1/2}|\bT|^{1/2}}
		\exp\left\{-\frac{1}{2\delta^2}
		\left[1-\frac{1}{\delta^2}\bgamma^\top\bT^{-1}\bgamma\right](v-\mu)^2\right\},
		\ee
		and so
		\be
		V &\sim& \N\left(\mu, \frac{\delta^2}{1-(\bgamma^\top\bT^{-1}\bgamma)/\delta^2}\right).
		\ee
	\end{proof}
	
\end{lemma}

\begin{proof} [Proof of Proposition 3.1.]
	Let $(X_1,X_2,\dots,X_q) \g \bSigma \sim(\bzero, \bSigma)$ and
	consider the do operator $\textnormal{do}(X_s=\tilde{x})$ where $s\in\{2,\dots,q\}$ is the intervened node.
	The post-intervention distribution of $X_1$ is given by (Section 3)
	\be
	f(x_1\g \textnormal{do}(X_s=\tilde x), \bSigma) = \int \N(x_1\g\gamma_s \tilde x+\bgamma^\top\bx_{\pa(s)},\bSigma)\cdot \N(\bx_{\pa(s)}\g\bzero,\bSigma_{\pa(s),\pa(s)}) \, d\bx_{\pa(s)},
	\ee
	where
	\be
	X_1\g \bX_{\pa(s)}=\bx_{\pa(s)} &\sim&\N(\gamma_s \tilde x+\bgamma^\top\bx_{\pa(s)},\delta^2), \\
	\bX_{\pa(s)} &\sim& \N(\bzero, \bSigma_{\pa(s),\pa(s)}).
	\ee
	Applying  Lemma \ref{corollary_2} with
	$V = X_1$ and $\boldsymbol{U} = \bX_{\pa(s)}$ we obtain
	\be
	X_1\g \textnormal{do}(X_s=\tilde x), \bSigma &=& \N\left(\gamma_s \tilde x, \frac{\delta^2}{1-(\bgamma^\top\bT^{-1}\bgamma)/\delta^2}\right),
	\ee
	where
	\begin{eqnarray}
	\delta^2 &=& \bSigma_{1\g\fa(s)}, \nonumber \\
	(\gamma_s,\bgamma^\top)^\top &=& \bSigma_{1,\fa(s)}\left(\bSigma_{\fa(s),\fa(s)}\right)^{-1}, \nonumber  \\
	\bT &=& \left(\bSigma_{\pa(s),\pa(s)}\right)^{-1} + \frac{1}{\delta^2}\bgamma\bgamma^\top. \nonumber
	\end{eqnarray}
\end{proof}

\section{PAS algorithm}
\label{subsec:Appendix:PAS}

We first summarize the main features of a \emph{Partial Analytic Structure} (PAS) algorithm \citep{Godsill:2012, wang:li:2012}.
Let $X_1,\dots,X_q$ be a collection of variables, $\bX$ a $(n,q)$ data matrix collecting $n$ i.i.d. multivariate observations from $X_1,\dots,X_q$. Consider $K$ distinct models, $\M_1,\dots, \M_K$, each one indexed by a parameter $\btheta_k\in\bTheta_k$ and assume model uncertainty, so that the true data generating model is one of the $K$ models. Under each model $\M_k$ the likelihood function is $f(\bX\g\btheta_k,\M_k)$.

Consider now two models $\M_h$, $\M_k$, $h,k\in\{1,\dots,K\}$, $h\ne k$, with parameters $\btheta_h, \btheta_k$ and let $(\btheta_h)_u$, $(\btheta_k)_u$ be two sub-vector of $\btheta_h, \btheta_k$ respectively.
A general PAS algorithm relies on the following two assumptions:
\begin{enumerate}
	\item the full conditional distribution of $(\btheta_k)_u$, $p\left((\btheta_k)_u\g(\btheta_k)_{-u},\M_k,\bX\right)$ is available in closed form;
	\item in $\M_h$ there exists an ``equivalent" set of parameters $(\btheta_h)_u$ with same dimension of $(\btheta_k)_u$;
\end{enumerate}
see also \citet[Section 5.2]{wang:li:2012}.
A PAS reversible jump algorithm adopts a proposal distribution which sets $(\btheta_h)_{-u} = (\btheta_k)_{-u}$, draws $\M_k$ from $q(\M_k\g\M_h)$ and $(\btheta_k)_u$ from $p\left((\btheta_k)_u\g(\btheta_k)_{-u},\M_k,\bX\right)$.
Specifically, the update of model $\M_k$ and model parameter $\btheta_k$ is performed in two steps.
The first step concerns the model update and can be summarized as follows:
\begin{enumerate}
	\item propose $\M_k\sim q(\M_k\g\M_h)$ and set $(\btheta_h)_{-u} = (\btheta_k)_{-u}$;
	\item accept $\M_k$ with probability $\alpha=\min\{1,r_{k}\}$,
	\ben
	\label{eq:PAS:accept}
	r_k = \frac{p(\M_k\g(\btheta_k)_{-u},\bX)}{p(\M_h\g(\btheta_h)_{-u},\bX)}
	\frac{q(\M_h\g\M_k)}{q(\M_k\g\M_h)},
	\een
	where
	\ben
	p(\M_k\g(\btheta_k)_{-u},\bX) = \int p(\M_k, (\btheta_k)_u \g (\btheta_k)_{-u},\bX)\, d(\btheta_k)_u
	\een
	\item if $\M_k$ is accepted, generate
	$(\btheta_k)_u \sim p\left((\btheta_k)_u\g(\btheta_k)_{-u},\M_k,\bX\right)$, \\otherwise
	$(\btheta_h)_u \sim p\left((\btheta_h)_u\g(\btheta_h)_{-u},\M_h,\bX\right)$.
\end{enumerate}
In the second step we then update parameter $\btheta_k$ (if $\M_k$ is accepted in the model update step) or $\btheta_h$ (otherwise) from its full conditional distribution using standard MCMC steps. Notice that in the case of Gaussian undirected graphical models with G-Wishart priors this requires specific MCMC techniques to avoid the computation of prior normalizing constants which are not available in closed form; see \cite[Section 2.4]{wang:li:2012}.

\vspace{0.5cm}

We now apply the PAS algorithm to our DAG setting.
We start defining a suitable proposal distribution, $q(\D'\g\D)$ on the DAG space.
To this end we consider three types of operators that locally modify a given DAG $\D$: insert a directed edge (InsertD $a\rightarrow b$ for short), delete a directed edge (DeleteD $a\rightarrow b$) and reverse a directed edge (ReverseD $a\rightarrow b$).
For each $\D\in\mathcal{S}_q$, being $\mathcal{S}_q$ the set of all DAGs on $q$ nodes, we then construct the set of \textit{valid} operators $\mathcal{O}_{\D}$, that is operators whose resulting graph is a DAG.
Therefore, given the current $\D$ we propose $\D'$ by uniformly sampling a DAG in $\mathcal{O}_\D$.
Because there is a one-to-one correspondence between each operator and resulting DAG $\D'$,
the probability of transition is given by
$q(\D'\g\D) = 1/|\mathcal{O}_{\D}|$.
Such a proposal can be easily adapted to account for sparsity constraints on the DAG space, for instance by fixing a maximum number of edges and limiting the model space to those DAGs having at most a given number of edges, typically a small multiple of the number of nodes.

Because of the structure of our proposal, at each step of our MCMC algorithm we will need to compare two DAGs $\D$, $\D'$ which differ by one edge only. Notice that the ReverseD $a\rightarrow b$ operator can be also brought back to the same case since is equivalent to the consecutive application of the operators DeleteD $a\rightarrow b$ and InsertD $b\rightarrow a$.
Therefore, consider two DAGs $\D=(V,E)$, $\D'=(V,E')$ such that $E'=E\setminus\{(h,j)\}$.
Notice that if a parent ordering is valid for $\D$, it is also valid for $\D'$, and we adopt this convention to simplify the analysis.
At this stage, for better clarity of notation, we index each parameter with its own DAG-model and write accordingly
$(\bD^{\D}, \bL^{\D})$ and $(\bD^{\D'}, \bL^{\D'})$ and $\bOmega^{\D}$ and $\bOmega^{\D'}$ when interest centers on the covariance matrix.
Notice that the Cholesky parameters under the two DAGs differ only with regard to their $j$-th component
$((\sigma_j^{\D })^2, \bL^{\D}_{\prec j\,]})$, and $((\sigma_j^{\D'})^2, \bL^{\D'}_{\prec j\,]})$ respectively.
Moreover the remaining  parameters
$\{ (\sigma_r^{\D})^2, \bL^{\D}_{\prec r\,]}; \, r \neq j \}$ and
$\{ (\sigma_r^{\D'})^2, \bL^{\D'}_{\prec r\,]}; \, r \neq j \}$
are componentwise \emph{equivalent} between the two graphs because they refer to structurally equivalent  conditional models; see
(6).
This is crucial for the correct application of the PAS algorithm.

The acceptance probability for $\D'$  under a PAS algorithm is given by
$\alpha_{\D'}=\min\{1;r_{\D'}\}$
where
\ben
\label{eq:ratio:PAS:DAG}
r_{\D'}
&=&\frac{p(\D'\g\bD^{\D'}\setminus (\sigma_j^{\D'})^2,\bL^{\D'}\setminus \bL^{\D'}_{\prec j\,]},\bX)}
{p(\D\g\bD^{\D}\setminus (\sigma_j^{\D})^2,\bL^{\D}\setminus \bL^{\D}_{\prec j\,]},\bX)}
\cdot
\frac{q(\D\g\D')}{q(\D'\g\D)}
\nonumber \\
&=&
\frac{p(\bX,\bD^{\D'}\setminus (\sigma_j^{\D'})^2,\bL^{\D'}\setminus \bL^{\D'}_{\prec j\,]}\g \D')}
{p(\bX,\bD^{\D}\setminus (\sigma_j^{\D})^2,\bL^{\D}\setminus \bL^{\D}_{\prec j\,]}\g \D)}
\cdot\frac{p(\D')}{p(\D)}
\cdot\frac{q(\D\g\D')}{q(\D'\g\D)}.
\een
Therefore we require to evaluate for DAG $\D$
\be
p(\bX,\bD\setminus \sigma_j^2,\bL\setminus \bL_{\prec j\,]}\g \D)
&=&
\int_0^{\infty}\int_{\Re^{|\pa(j)|}}
p(\bX\g\bD,\bL,\D)p(\bD,\bL\g\D) \, d \bL_{\prec j\,]} d\sigma_j^2
\ee
(and similarly for $\D'$) where
we removed for simplicity the super-script $\D$ from the Cholesky parameters, now emphasizing the dependence on $\D$ though the conditioning sets.
Moreover, because of the likelihood and prior factorization in (8) and (16) we can write
\ben
\label{eq:PAS:joint:DAG}
p(\bX,\bD\setminus \sigma_j^2,\bL\setminus \bL_{\prec j\,]}\g \D)
&=&
\prod_{r\ne j}f(\bX_r\g\bX_{\pa(r)},\sigma_r^2,\bL_{\prec r\,]},\D)p(\bL_{\prec r\,]}\g\sigma_r^2,\D)p(\sigma_r^2\g\D) \nonumber \\
&\cdot&
\int_0^{\infty}\int_{\Re^{|\pa(j)|}}
f(\bX_j\g\bX_{\pa(j)},\sigma_j^2,\bL_{\prec j\,]},\D) \\
&\cdot&
p(\bL_{\prec j\,]}\g\sigma_j^2,\D)p(\sigma_j^2\g\D)
\, d \bL_{\prec j\,]} d\sigma_j^2. \nonumber
\een
In particular, because of conjugacy of the Normal density with the Normal-Inverse-Gamma prior, the integral in \eqref{eq:PAS:joint:DAG} can be obtained in closed form as
\ben
\label{eq:marg:like:j:suppl}
m_{}(\bX_j\g\bX_{\pa_{\D}(j)}, \D) =
(2\pi)^{-\frac{n}{2}}
\frac{\big|\bT_j\big|^{1/2}}
{\big|\bar{\bT}_j\big|^{1/2}}\cdot
\frac{\Gamma\left(a_j^*+\frac{n}{2}\right)}{\Gamma\left(a_j^*\right)}
\left[\frac{1}{2}g\right]^{a_j^*}
\left[\frac{1}{2}\big(g+\bX_j^\top\bX_j - \hat{\bL}_j^\top\bar{\bT}_j\hat{\bL}_j\big)\right]^{-(a_j^*+n/2)}
\een
where
\be
\bT_j &=& g\bI_{|\pa_{\D}(j)|} \\
\bar{\bT}_j &=& g\bI_{|\pa_{\D}(j)|}+\bX_{\pa_{\D}(j)}^\top\bX_{\pa_{\D}(j)} \\
\hat{\bL}_j
&=& \big(g\bI_{|\pa_{\D}(j)|}+\bX_{\pa_{\D}(j)}^\top\bX_{\pa_{\D}(j)}\big)^{-1}\bX_{\pa_{\D}(j)}^{\top}\bX_j,
\ee
and $a_j^*=\frac{a_j}{2}-\frac{|\pa_{\D}(j)|}{2}-1$. For $j=1$, because we fixed $\sigma_1^2=1$ we instead obtain
\ben
m_{}(\bX_1\g\bX_{\pa_{\D}(1)}, \D) &=&
(2\pi)^{-\frac{n}{2}} \frac{\big|\bT_1\big|^{1/2}}
{\big|\bar{\bT}_1\big|^{1/2}}\cdot
\exp\left\{-\frac{1}{2}\big(\bX_1^\top\bX_1+
\hat{\bL}_1^\top\bar{\bT}_1\hat{\bL}_1\big)\right\}.
\een
Therefore, the PAS ratio in \eqref{eq:ratio:PAS:DAG} reduces to
\ben
r_{\D'}&=&
\frac{m_{}(\bX_j\g\bX_{\pa_{\D'}(j)}, \D')}
{m_{}(\bX_j\g\bX_{\pa_{\D}(j)}, \D)}
\cdot\frac{p(\D')}{p(\D)}
\cdot\frac{q(\D\g\D')}{q(\D'\g\D)}.
\een

\section{Proof of Proposition 4.1}

In this section we prove that the posterior distribution $p(\bD,\bL,\D,\theta_0,\bX_1\g \by, \bX_{-1})$ is proper. To this end we factorize it as
\be
p(\bD,\bL,\D,\theta_0,\bX_1 \g \by, \bX_{-1}) &\propto&
p(\theta_0\g \by, \bX_{-1})
p(\bX_1 \g \theta_0,\by, \bX_{-1}) \\
&\cdot& p(\bD,\bL \g \theta_0,\bX_1, \by, \bX_{-1})
p(\D \g \bD,\bL,\theta_0,\bX_1, \by, \bX_{-1}).
\ee
Also, because $\by$ is deterministically set once $\bX_1$ and $\theta_0$ are given, the latter simplifies to
\be
p(\bD,\bL,\D,\theta_0,\bX_1 \g \by, \bX_{-1}) &\propto&
p(\theta_0\g \by, \bX_{-1})
p(\bX_1 \g \theta_0,\by, \bX_{-1}) \\
&\cdot& p(\bD,\bL \g \theta_0,\bX)
p(\D \g \bD,\bL,\theta_0,\bX),
\ee
where $\bX$ is the $(n,q)$ augmented data matrix, column binding of $\bX_1$ and $\bX_{-1}$.
To prove the propriety of the joint posterior we will show that each of the above terms corresponds to a proper distribution.

\subsection{Propriety of \boldmath$p(\theta_0\g y, X_{-1})$}
 
Frist notice that $p(\theta_0\g \by, \bX_{-1}) \propto p(\by,\bX_{-1}\g \theta_0)$ since $p(\theta_0)\propto 1$.
Remember the \textit{augmented} likelihood in Equation (9) of our manuscript
\be
f(\by,\bX\g \bD,\bL,\theta_0,\D) =
\prod_{j=1}^{q} d\,\N_n(\bX_j\g -\bX_{\pa(j)} \bL_{ \prec j\,]}, \sigma_j^2 \bI_n)
\cdot
\left\{
\prod_{i=1}^{n}
\mathbbm{1}(\theta_{y_i-1} < x_{i,1} \le \theta_{y_i})
\right\},
\ee
now emphasizing the dependence on DAG $\D$, where we set the $(j,j)$-element of $\bD$ equal to $\sigma_j^2$.
We first integrate out $\bX_1$ to obtain the likelihood
\be
f(\by,\bX_{-1}\g \bD,\bL,\theta_0, \D) &=&
\prod_{j=2}^{q} d\,\N_n(\bX_j\g -\bX_{\pa(j)} \bL_{ \prec j\,]}, \sigma_j^2 \bI_n) \\
&\cdot&
\int_{\Re^n}
d\,\N_n(\bX_1\g -\bX_{\pa(1)} \bL_{ \prec 1\,]}, \sigma_1^2 \bI_n)
\prod_{i=1}^{n}
\mathbbm{1}(\theta_{y_i-1} < x_{i,1} \le \theta_{y_i}) \,
d\bX_1 \\
&=&
f(\bX_{-1}\g\bD,\bL) \cdot
\int\limits_{O^n(\by,\theta_0)}
d\,\N_n(\bX_1\g -\bX_{\pa(1)} \bL_{ \prec 1\,]}, \sigma_1^2 \bI_n) \,
d\bX_1,
\ee
where $\bX_{-1}=(\bX_2,\dots,\bX_q)$, $O^n(\by,\theta_0)$ is the \textit{n}-dimensional orthant $\bigtimes\limits_{i=1}^n(\theta_{y_i-1}, \theta_{y_i}]$ and $\theta_{-1}=-\infty, \theta_{1}=+\infty$.
To obtain the marginal posterior of $\theta_0$ we 
first compute the \textit{integrated likelihood} for $\theta_0$ by 
integrating the likelihood w.r.t. $(\bD,\bL)$. Because of prior parameter independence across $(\sigma_j^2,\bL_{ \prec j\,]})$'s, we obtain
\be
f(\by,\bX_{-1}\g\theta_0) &=&
\int f(\by,\bX_{-1}\g \bD,\bL,\theta_0) \, p(\bD,\bL)\, d \bD \, d\bL \\
&=& \prod_{j=2}^{q} \left\{\, \int
d\,\N_n(\bX_j\g -\bX_{\pa(j)} \bL_{ \prec j\,]}, \sigma_j^2 \bI_n)
\,p(\sigma_j^2,\bL_{ \prec j\,]})
\, d\sigma_j^2 \, d\bL_{ \prec j\,]} \right\} \\
&\cdot&
\int\limits_{\Re^{|\pa(1)|}}
\int\limits_{O^n(\by,\theta_0)}
d\,\N_n(\bX_1\g -\bX_{\pa(1)} \bL_{ \prec 1\,]}, \sigma_1^2 \bI_n)
\, p(\bL_{ \prec 1\,]})\, d\bX_1 \, d\bL_{ \prec 1\,]},
\ee
since $\sigma_1^2=1$.
Notice that the first term in the integrated likelihood $f(\by,\bX_{-1}\g\theta_0)$ does not depend on $\theta_0$. We then write
\be
f(\by,\bX_{-1}\g\theta_0) =
K(\bX_{-1}) \cdot
\int\limits_{O^n(\by,\theta_0)}
\int\limits_{\Re^{|\pa(1)|}}
d\,\N_n(\bX_1\g -\bX_{\pa(1)} \bL_{ \prec 1\,]}, \sigma_1^2 \bI_n)
\, p(\bL_{ \prec 1\,]}) \, d\bL_{ \prec 1\,]} \, d\bX_1,
\ee
upon interchanging the order of integration.
Consider now the inner integral where $p(\bL_{ \prec 1\,]})= d\,\N_{|\pa(1)|}(\bzero,g^{-1}\bI_{\pa(1)})$; see also Equation (15) in our manuscript.
Because of conjugacy with the normal density $d\,\N_n(\bX_1\g\cdot)$ we obtain
\be
\int\limits_{\Re^{|\pa(1)|}}
d\,\N_n(\bX_1\g -\bX_{\pa(1)} \bL_{ \prec 1\,]}, \sigma_1^2 \bI_n)
\, p(\bL_{ \prec 1\,]}) \, d\bL_{ \prec 1\,]} =
d\,\N_n(\bX_1\g\bzero,\bSigma_1),
\ee
where $|\pa(1)|$ is the number of parents,
$
\bSigma_1 = g^{-1} \bX_{\pa(1)}\bX_{\pa(1)}^\top + \bI_n.
$
The integrated likelihood thus becomes
\ben
\label{eq:integrated:likelihood}
f(\by,\bX_{-1}\g\theta_0) =
K(\bX_{-1}) \cdot
\int\limits_{O^n(\by,\theta_0)}
d\,\N_n(\bX_1\g\bzero,\bSigma_1)
\, d\bX_1.
\een
Since we assumed $p(\theta_0)\propto 1$, the marginal posterior of $\theta_0$ is proportional to \eqref{eq:integrated:likelihood} and also
\be
\label{eq:marginal:posterior}
p(\theta_0\g\by,\bX_{-1}) 
&\propto& \int\limits_{O^n(\by,\theta_0)}
d\,\N_n(\bX_1\g\bzero,\bSigma_1)\, d\bX_1 \\
&\coloneqq& I(\theta_0,n),
\ee
where we emphasize the dependence on $n$. Therefore, to verify the propriety of $p(\theta_0\g\by,\bX_{-1})$ we need to prove that $I(\theta_0,n)$ is integrable over $\theta_0\in(-\infty,\infty)$.

\vspace{0.5cm}

\noindent Let now for simplicity $\bX_1 = \bU=(U_1,\dots,U_n)^\top$ and recall that
\be
U_i \in (-\infty,\theta_0] \iff U_i \le \theta_0 \iff U_i - \theta_0 \le 0, \\
U_i \in (\theta_0,\infty) \iff U_i > \theta_0 \iff \theta_0 - U_i < 0.
\ee
Hence, if we let
\be
U_i^* =
\begin{cases}
	\,\, U_i-\theta_0 & \text{if $y_i=0$},\\
	\,\, \theta_0-U_i & \text{if $y_i=1$},
\end{cases}
\ee
we obtain that $\bU^* \sim \N_n(\bmu^*,\bSigma^*)$, where $\bmu^*=(\mu^*_1,\dots,\mu^*_n)^\top$,
\be
\mu_i^* =
\begin{cases}
	\,\, -\theta_0 & \text{if $y_i=0$},\\
	\,\, \theta_0 & \text{if $y_i=1$},
\end{cases} \quad \quad
\bSigma^*(h,k) =
\begin{cases}
	\,\, \bSigma_1(h,k) & \text{if $y_h = y_k$},\\
	\,\, -\bSigma_1(h,k) & \text{if $y_h \ne y_k$},
\end{cases}
\ee
and $\bSigma^*(h,k)$ denotes the element at position $(h,k)$ in $\bSigma^*$. Therefore, we can write
\ben
\label{eq:integral:n}
I(\theta_0,n) &=&
\int\limits_{0}^{\infty}
\cdots
\int\limits_{0}^{\infty}
d\,\N_n(\bU^*\g\bmu^*,\bSigma^*) \, d U_1^* \cdots \, d U_n^*.
\een

\vspace{0.5cm}
\noindent If $n=1$, then \eqref{eq:integral:n} is  not integrable so that the posterior of $\theta_0$ improper.
To see why, notice that
\be
I(\theta_0,1)
&=&
\int\limits_{0}^{\infty}
d\,\N(U_1^*\g \mu^*,\sigma^{2*}) \, d U_1^* \\
&=&
\begin{cases}
	\,\, \Phi\left(\frac{\theta_0}{\sigma^{2*}}\right)\ & \text{if $y_i=0$},\\
	\,\, \Phi\left(-\frac{\theta_0}{\sigma^{2*}}\right) & \text{if $y_i=1$},
\end{cases}
\ee
which  is not integrable over $(-\infty,\infty)$ in either case.

\vspace{0.4cm}
\noindent Consider now $n=2$ and assume that the two observations have different values for $Y$, say $y_1=1$ and $y_2=0$.
We then obtain
\begin{eqnarray*}
	\begin{pmatrix}
		U_2^*\\
		U_2^*
	\end{pmatrix} & \sim & \N\left[\left(\begin{array}{c}
		\theta_0\\
		-\theta_0
	\end{array}\right),\left(\begin{array}{cc}
		\sigma_1^2 & \rho\sigma_1\sigma_2 \\
		\rho\sigma_1\sigma_2 & \sigma_2^2
	\end{array}\right)\right],
\end{eqnarray*}
where $\rho=\textnormal{Corr}(U_1^*,U_2^*)$ and
\be
\label{eq:integral}
I(\theta_0,2) &=&
\int\limits_{0}^{\infty}
\int\limits_{0}^{\infty}
d\,\N_n(\bU^*\g\bmu^*,\bSigma^*) \, dU_1^* \, dU_2^* \\
&=&
P\left\{U_1^*>0,U_2^*>0\right\}.
\ee 
The posterior for $\theta_0$ is  proper
if and only if $\int_{-\infty}^{\infty}I(\theta_0,2)\, d\theta_0 < \infty$.
We will show this result by providing an upper bound
$I(\theta_0,2) \leq G(\theta_0,2)$ with $G(\theta_0,2)$ integrable over the real line.
To this end, we first standardize $U_1^*$ and $U_2^*$
\be
\label{eq:U:standardized}
V_1^*=
\frac{U_1^*-\theta_0}{\sigma_1} \\
V_2^*=\frac{U_2^*+\theta_0}{\sigma_2},
\ee
so that
\begin{eqnarray*}
	\begin{pmatrix}
		V_2^*\\
		V_2^*
	\end{pmatrix} & \sim & \N\left[\left(\begin{array}{c}
		0 \\
		0
	\end{array}\right),\left(\begin{array}{cc}
		1 & \rho \\
		\rho\ & 1
	\end{array}\right)\right]
\end{eqnarray*}
and
\ben
\label{eq:integral:U:V}
P\left\{U_1^*>0,U_2^*>0\right\} = P\left\{V_1^*>-\frac{\theta_0}{\sigma_1},V_2^*>\frac{\theta_0}{\sigma_2}\right\}.
\een
We now distinguish  two cases: $\rho\ge 0$ and  $\rho < 0$.

\paragraph{Case 1}
Consider first $0 \le \rho < 1 $. Applying Equation (C.8) in \citet{Beck:Melchers:2017} to the right-hand-side of
\eqref{eq:integral:U:V}
we obtain
\be
P\left\{U_1^*>0,U_2^*>0\right\}
\le
\Phi(-b)\Phi\left(\frac{\theta_0}{\sigma_1}\right) + \Phi(-a)\Phi\left(-\frac{\theta_0}{\sigma_2}\right)
\coloneqq H(\theta_0,2),
\ee
where
\be
a=\frac{-\theta_0\left[\frac{1}{\sigma_1}+\frac{\rho}{\sigma_2}\right]}{(1-\rho^2)^{1/2}}, \quad \quad
b=\frac{\theta_0\left[\frac{1}{\sigma_2}+\frac{\rho}{\sigma_1}\right]}{(1-\rho^2)^{1/2}},
\ee
and $\Phi(t)$ is the c.d.f. of a standard normal distribution evaluated at $t$.
It follows that $H(\theta_0,2)$ is continuous, bounded and
\be
\lim_{\theta_0\rightarrow -\infty} H(\theta_0,2) = 0, \quad \quad
\lim_{\theta_0\rightarrow \infty} H(\theta_0,2) = 0.
\ee
Now we verify integrability of $H(\theta_0,2)$ in a right and left neighborhood of $-\infty$ and $\infty$ respectively.

\vspace{0.2cm}
Consider first $\theta_0\rightarrow-\infty$ and notice that
\be
H(\theta_0,2) &\le&
\Phi\left(\frac{\theta_0}{\sigma_1}\right) +
\Phi\left(-a\right) \\
&=&
\Phi\left(\theta_0 A\right) +
\Phi\left(\theta_0 B\right) \\
&=&
P\left\{Z\ge-\theta_0 A\right\} + P\left\{Z \ge -\theta_0 B\right\}.
\ee
where $Z\sim\N(0,1)$ and
\be
A = \frac{1}{\sigma_1}>0 \quad \quad
B = \frac{\frac{1}{\sigma_1}+\frac{\rho}{\sigma_2}}{(1-\rho^2)^{1/2}} > 0.
\ee
Because $\theta_0 \rightarrow -\infty$,  we have $-\theta_0 A>0, -\theta_0 B > 0$. Applying the following inequality for the upper tail of $Z$  
\be
P\{Z > z\} \le \frac{\exp\{-z^2/2\}}{z\sqrt{2\pi}}, \, z>0, 
\ee
one gets
\be
P\left\{Z\ge-\theta_0 A\right\} + P\left\{Z \ge -\theta_0 B\right\}
&\le&
\frac{\exp\{-\theta_0^2 A^2/2\}}{-\theta_0 A \sqrt{2\pi}} +
\frac{\exp\{-\theta_0^2 B^2/2\}}{-\theta_0 B \sqrt{2\pi}} \\
&\le&
\frac{\exp\{-\theta_0^2 A^2/2\}}{A \sqrt{2\pi}} +
\frac{\exp\{-\theta_0^2 B^2/2\}}{B \sqrt{2\pi}} \\
&\coloneqq&
G(\theta_0,2).
\ee
Clearly $\int_{-\infty}^0 G(\theta_0,2) d\theta_0 < \infty$, whence  
$I(\theta_0,2)$ is integrable  in a right neighborhood of $-\infty$.

\vspace{0.2cm}
Consider now $\theta_0\rightarrow\infty$.  Similarly as before notice that
\be
H(\theta_0,2) &\le&
\Phi\left(-b\right) +
\Phi\left(-\frac{\theta_0}{\sigma_2}\right) \\
&=&
\Phi\left(-\theta_0 C\right) +
\Phi\left(-\theta_0 D\right) \\
&=&
P\left\{Z\ge \theta_0 C\right\} + P\left\{Z \ge \theta_0 D\right\}.
\ee
where
\be
C = \frac{\frac{1}{\sigma_2}+\frac{\rho}{\sigma_1}}{(1-\rho^2)^{1/2}} > 0, \quad \quad
D = \frac{1}{\sigma_2}>0
\ee
It follows that an upper bound for $I(\theta_0,2)$ is now
\be
G(\theta_0,2) =
\frac{\exp\{-\theta_0^2 C^2/2\}}{C \sqrt{2\pi}} +
\frac{\exp\{-\theta_0^2 D^2/2\}}{D \sqrt{2\pi}}.
\ee
Thus for $ \rho \geq 0 $ the posterior for $\theta_0$ is proper.

\paragraph{Case 2}

Consider now the case $-1<\rho<0$.

We rewrite \eqref{eq:integral:U:V} using  $P_{\rho<0}\left\{\cdot \right\}$ to indicate that the probability is evaluated w.r.t. the joint distribution when the correlation coefficient is less than  zero, and use $P_{\rho \ge 0}\left\{\cdot \right\}$ otherwise.
Letting $v_1=-\frac{\theta_0}{\sigma_1}, v_2=\frac{\theta_0}{\sigma_2}$ we have
\be
P_{\rho<0}\left\{U_1^*>0,U_2^*>0\right\} &=&
P_{\rho<0}\left\{V_1^*>v_1,V_2^*>v_2\right\} \\
&=&
P_{\rho<0}\left\{V_1^*\le v_1,V_2^* \le v_2\right\} +
P\left\{V_1^*> v_1\right\} +
P\left\{V_2^*> v_2\right\} - 1 \\
&\le&
P_{\rho \ge 0}\left\{V_1^*\le v_1,V_2^* \le v_2\right\} +
P\left\{V_1^*> v_1\right\} +
P\left\{V_2^*> v_2\right\} - 1 \\
&=&
P_{\rho \ge 0}\left\{V_1^*> v_1,V_2^* > v_2\right\} \\
&=&
P_{\rho \ge 0}\left\{U_1^*> 0, U_2^* > 0 \right\};
\ee
see Equation (C.16) in \citet{Beck:Melchers:2017} for the inequality in step 3.

It follows that $I(\theta_0,2)$ for $\-1<\rho<0$ is bounded above by the corresponding function for the case $0 \le \rho <1$.
Hence $I(\theta_0,2)$ is integrable also for $\rho<0$ and propriety of the posterior of $\theta_0$ holds in  this case too.

\vspace{0.5cm}

Having established  that $p(\theta_0\g\by,\bX_{-1})$ is proper when $n=2$ and $y_1=1, y_2=0$, it follows that propriety will hold  for any sample size $n\ge2$ provided that there exist at least two observations with distinct values for $Y$. To see why, assume without loss of generality that the first two observations have  distinct values;  then
\be
p\left(\theta_0\g\by^{(1:n)}, \bX_{-1}^{(1:n)}\right) \propto  
p\left(\theta_0\g\by^{(1:2)}, \bX_{-1}^{(1:2)}\right) \cdot
f\left(\by^{(3:n)}, \bX_{-1}^{(3:n)} \g \by^{(1:2)}, \bX_{-1}^{(1:2)}, \theta_0 \right),
\ee
which is proper because 
$p\left(\theta_0\g\by^{(1:2)}, \bX_{-1}^{(1:2)}\right)$  is proper and 
the conditional integrated likelihood $f(\cdot\g\cdot)$ is not degenerate.

\vspace{0.8cm}

\noindent
Results presented in the next sections rely on the following proposition.

\begin{prop}
	\label{prop:integrated:like:proper}
	Let $p(\bX\g \boldsymbol{\vartheta}, \boldsymbol{\lambda})$ be a proper statistical model for the data $\bX$ where $\boldsymbol{\vartheta}$, $\boldsymbol{\lambda}$ are continuous parameters with joint prior
	$p(\boldsymbol{\vartheta}, \boldsymbol{\lambda})=
	p(\boldsymbol{\lambda}\g\boldsymbol{\vartheta})p(\boldsymbol{\vartheta})$.
	If $p(\boldsymbol{\lambda}\g\boldsymbol{\vartheta})$ is proper, then
	\be
	p(\bX \g \boldsymbol{\vartheta}) =
	\int p(\bX\g \boldsymbol{\vartheta}, \boldsymbol{\lambda})
	p(\boldsymbol{\lambda}\g\boldsymbol{\vartheta}) \, d\boldsymbol{\lambda}
	\ee
	is also a proper statistical model.
	If viewed as a function of $\boldsymbol{\vartheta}$, $p(\bX \g \boldsymbol{\vartheta})$ is called \textit{integrated likelihood}.
	
	\begin{proof}
		The proof is immediate if one can interchange the order of integration between $\bX$ and $\boldsymbol{\lambda}$.
	\end{proof}
	\noindent
	In addition, it follows that if $p(\boldsymbol{\vartheta})$ is also proper, then the posterior of $\boldsymbol{\vartheta}$, $p(\boldsymbol{\vartheta}\g\bX)$, will also be proper.
\end{prop}

\subsection{Propriety of \boldmath $p(X_1 \g \theta_0,y, X_{-1})$}

We can first write
\be
p(\bX_1 \g \theta_0, \by, \bX_{-1}) &\propto&
p(\bX_{-1}\g \by,\theta_0,\bX_1)
p(\by\g \theta_0,\bX_1)
p(\bX_1\g\theta_0) \\
&=&
p(\bX_{-1}\g \theta_0,\bX_1)
p(\by\g \theta_0,\bX_1)
p(\bX_1\g\theta_0)
\ee
since $p(\bX_{-1}\g \by,\theta_0,\bX_1) = p(\bX_{-1}\g \theta_0,\bX_1)$ and $p(\theta_0)\propto 1$.

\vspace{0.4cm}
The first term can be written as
\be
p(\bX_{-1}\g \theta_0,\bX_1) =
\sum_{\D}\left\{
\int \int p(\bX_{-1}\g \bD, \bL, \D,\theta_0,\bX_1)
p(\bD,\bL\g\D, \theta_0,\bX_1)
\, d\bD \, d\bL
\right\}
p(\D\g\theta_0,\bX_1).
\ee
Now notice that
\be
p(\bD,\bL\g\D, \theta_0,\bX_1)
&\propto&
p(\bX_1\g\bD,\bL,\D,\theta_0) p(\bD,\bL\g \D,\theta_0) \\
&\propto&
p(\bX_1\g\bD,\bL,\D,\theta_0) p(\bD,\bL\g \D),
\ee
and that we can write
\be
p(\bX_1\g\bD,\bL,\D,\theta_0)
&=&
\sum_{\by\in\{0,1\}^n}
\left\{
\int
f(\by,\bX\g \bD,\bL,\theta_0)
\, d\bX_{-1}
\right\} \\
&=&
\sum_{\by\in\{0,1\}^n}
\left\{
\int
f(\bX\g\bD,\bL,\D)
\, d \bX_{-1}
\prod_{i=1}^{n}
\mathbbm{1}(\theta_{y_i-1} < x_{i,1} \le \theta_{y_i})
\right\} \\
&=&
d\N_n(\bX_1\g\bzero,\tau_1^2\bI_n)
\sum_{\by\in\{0,1\}^n}
\left\{
\prod_{i=1}^{n}
\mathbbm{1}(\theta_{y_i-1} < x_{i,1} \le \theta_{y_i})
\right\} \\
&=&
d\N_n(\bX_1\g\bzero,\tau_1^2\bI_n),
\ee
being $\tau_1^2=[(\bL^\top)^{-1}\bD \bL^{-1}]_{11}$, because for each fixed $\bX_1=(x_{1,1},\dots,x_{n,1})^\top$ only one of the $2^n$ product of indicators will hold true and therefore $\sum_{y\in\{0,1\}^n}\{\dots\}=1$.
Hence, $p(\bX_1\g\bD,\bL,\D,\theta_0)$ corresponds to a (proper) multivariate normal density.
Also, $p(\bD,\bL\g\D)$ is a proper (prior) distribution by assumption.
It follows that $p(\bX_{-1}\g \theta_0,\bX_1)$ is an integrated likelihood because it is obtained with proper priors $p(\bD,\bL\g\D, \theta_0,\bX_1)$ and $p(\D\g \theta_0, \bX_1)$.

\vspace{0.3cm}
Moreover, $p(\by\g\theta_0,\bX_1)$ is also proper because
\ben
p(\by\g\theta_0,\bX_1)=
\begin{cases}
	1 & \text{if $y_i = \mathbbm{1}(x_{i,1}>\theta_0)$ for each $i=1,\dots,n$},\\
	0
	& \text{otherwise}.
\end{cases}
\een

\vspace{0.3cm}
Finally we have
\be
p(\bX_1\g\theta_0) &=&
\sum_{\D}\left\{
\int \int
p(\bX_1\g\bD,\bL,\D,\theta_0)
p(\bD,\bL\g\theta_0,\D)
\, d\bD \, d \bL
\right\}
p(\D\g\theta_0) \\
&=&
\sum_{\D}\left\{
\int
d\N_n(\bX_1\g\bzero,\tau_1^2\bI_n)
p(\tau_1^2\g\D)
\, d\tau_1^2
\right\}
p(\D),
\ee
which is also proper because the family of densities for $\bX_1$ is integrated w.r.t. proper priors
$p(\tau_1^2\g\D)$ (induced by the proper prior on $(\bD,\bL)$) and $p(\D)$.
\black

\subsection{Propriety of \boldmath $p(D,L \g \theta_0,X)$}

Consider now
\be
p(\bD,\bL \g \theta_0,\bX) =
\sum_{\D}
p(\bD,\bL \g \D, \theta_0,\bX)
p(\D\g \theta_0,\bX).
\ee
\vspace{0.3cm}
First notice that
\be
p(\bD,\bL \g \D, \theta_0,\bX)
&\propto&
p(\bX\g \bD,\bL,\D,\theta_0)
p(\bD,\bL\g\D,\theta_0) \\
&=&
p(\bX\g \bD,\bL,\D,\theta_0)
p(\bD,\bL\g\D),
\ee
where
\be
p(\bX\g\bD,\bL,\D,\theta_0)
&=&
\sum_{\by\in\{0,1\}^n}
p(\by,\bX\g \bD,\bL,\theta_0)
\\
&=&
\sum_{\by\in\{0,1\}^n}
\left\{
p(\bX\g\bD,\bL,\D)
\prod_{i=1}^{n}
\mathbbm{1}(\theta_{y_i-1} < x_{i,1} \le \theta_{y_i})
\right\} \\
&=&
p(\bX\g\bD,\bL,\D),
\ee
arguing as in Section 3.2. Therefore we can write $p(\bD,\bL\g\D,\theta_0,\bX) = p(\bD,\bL\g\D,\bX)$ which is a proper DAG Wishart distribution.

\vspace{0.3cm}
In addition,
$
p(\D\g\theta_0,\bX)
\propto
p(\bX\g\D,\theta_0) p(\D\g\theta_0) =
p(\bX\g\D,\theta_0) p(\D)
$
where
\be
p(\bX\g\theta_0,\D) &=&
\int \int
p(\bX\g\bD,\bL,\D,\theta_0) p(\bD,\bL\g\D,\theta_0)
\, d\bD \, d\bL \\
&=&
\int \int
p(\bX\g\bD,\bL,\D) p(\bD,\bL\g\D)
\, d\bD \, d\bL
\ee
is again an integrated (augmented) likelihood with a proper prior $p(\bD,\bL\g\D)$ and therefore proper.
Hence, $p(\D\g\theta_0,\bX)$ is proper according to Proposition \ref{prop:integrated:like:proper} because $p(\D)$ is also a proper prior. 

\vspace{0.3cm}
Therefore, $p(\bD,\bL \g \theta_0,\bX)$ is proper because it corresponds to a finite mixture of proper priors.

\subsection{Propriety of \boldmath $p(\D \g D,L,\theta_0,X)$}

Finally, we can write
\be
p(\D \g \bD,\bL,\theta_0,\bX) &\propto&
p(\bX\g\bD,\bL,\D,\theta_0)p(\D\g\theta_0) \\
&=&
p(\bX\g\bD,\bL,\D)p(\D),
\ee
which will be proper because $\D$ belongs to a finite space.

\section{MCMC convergence diagnostics}

In order to fix the numer of MCMC iterations in our simulation study we have run few pilot simulations that we used to perform diagnostics of convergence.
Specifically, for each pilot simulation we ran two independent chains of length $T_1$ and $T_2$. We fixed $T_1=25000$ and $T_2=50000$ in the $q=20$ setting, while $T_1=50000$ and $T_2=100000$ for $q=40$.

\begin{figure}
	\begin{center}
		\begin{tabular}{cc}
			\includegraphics[scale=0.22]{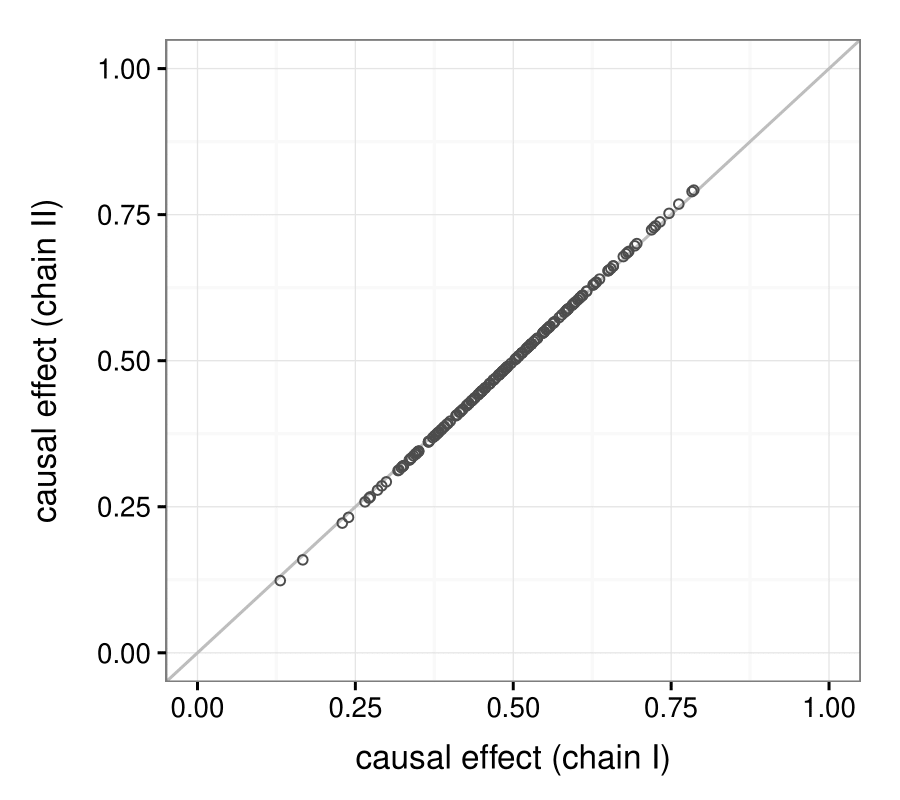} &
			\includegraphics[scale=0.22]{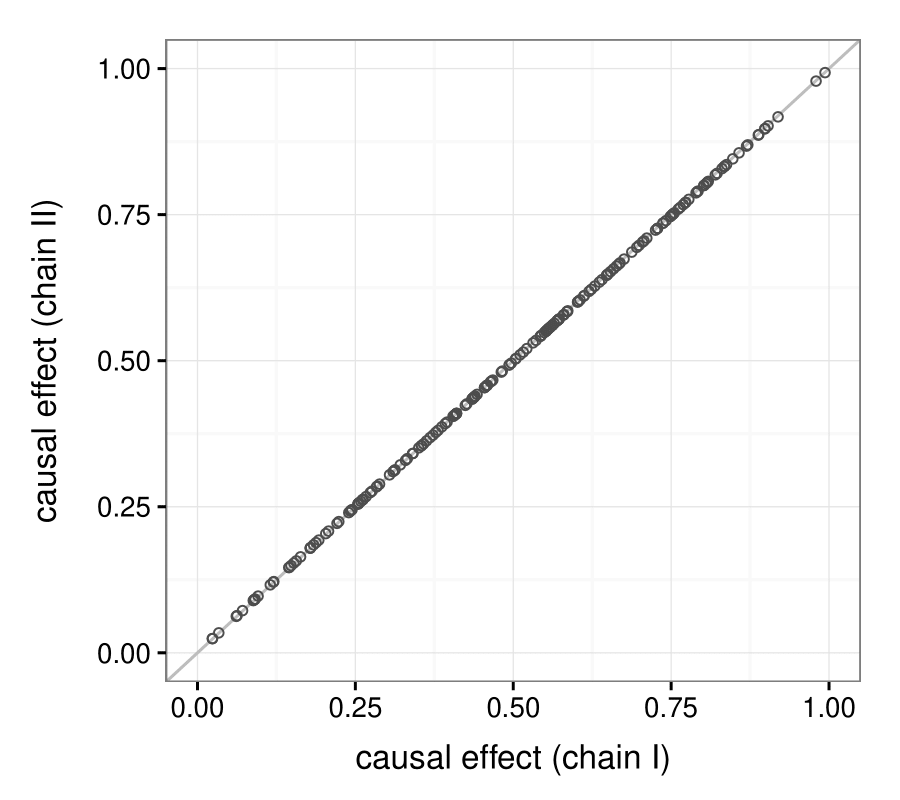} \\
			\includegraphics[scale=0.22]{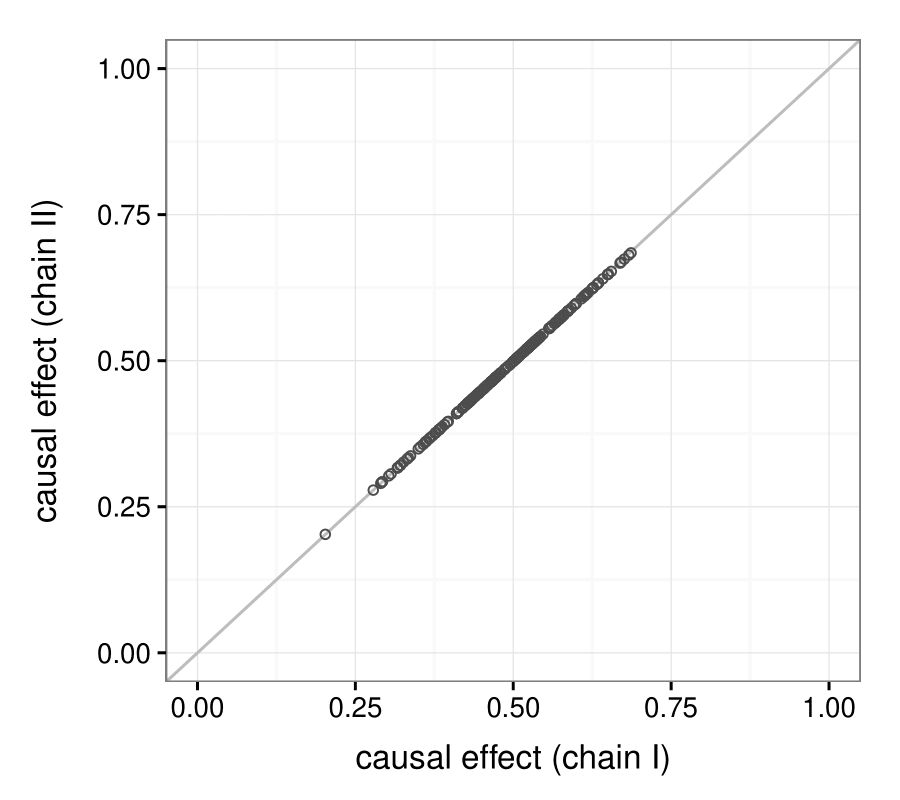} &
			\includegraphics[scale=0.22]{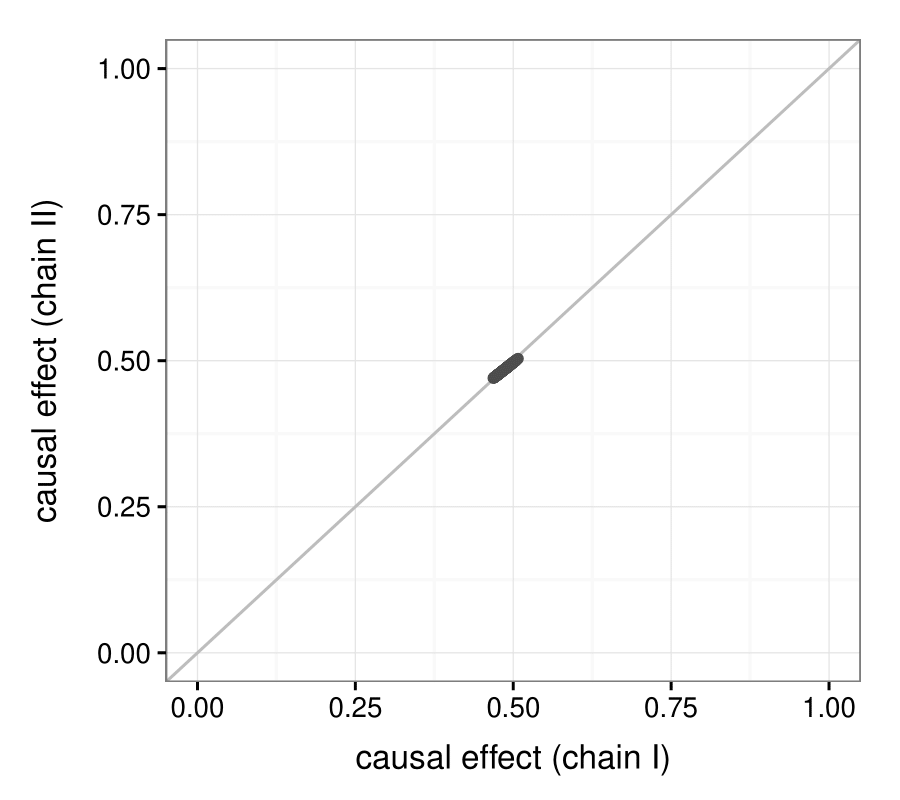}
		\end{tabular}
		\caption{\small Scatter plots of the BMA estimates for four randomly chosen intervened nodes, obtained from two chains of length $T_1=25000$ and $T_2=50000$ (pilot simulation for the $q=20$ setting).}
		\label{fig:conv_causal_effect}
	\end{center}
\end{figure}

We compare the BMA causal effect estimates (as defined in Equation (27) of our paper) obtained under the two independent chains.
Figure \ref{fig:conv_causal_effect} shows the scatter plots of the BMA estimates obtained from the two chains for four randomly chosen intervened nodes in one of the pilot simulations for $q=20$.
By inspection, one can see that the agreement between the results is highly satisfactory since points are clustered around the main diagonal of the plot.
Therefore, $T_1$ iterations are sufficient to reach convergence. Similar results, not reported for brevity, were obtained in the $q=40$ scenarios.


\section{Simulation results and comparisons}

\begin{figure}
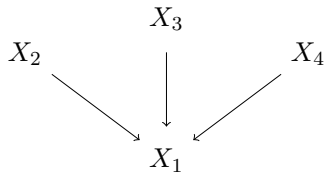

	\begin{center}
		\begin{tabular}{c}
			{\normalsize
				\tikz \graph [no placement, math nodes, nodes={circle}]
				{
					{X_2[x=0,y=-0.5], X_3[x=2,y=0], X_4[x=4,y=-0.5]}-> X_1[x=2,y=-2]};
			}
		\end{tabular}
	\end{center}
	\caption{\small An instance of \textit{naive} DAG with $q=4$ nodes used in the alternative method under comparison.}
	\label{fig:regr:dag}
\end{figure}

To emphasize the crucial role played by the set $fa(s)$ in causal effect estimation, we compare our method with an alternative setup which does not adjust for confounders.
Specifically, we perform inference of causal effects under a fixed \textit{naive} DAG wherein all variables, save for the response, are disjoint and each is a parent of the response; see Figure \ref{fig:regr:dag} for an example on $q=4$ nodes.
The resulting DAG depicts a situation similar to that of a standard (Bayesian) probit regression model, wherein the covariates are not fixed but rather random and jointly independent with marginal normal distributions, $X_j\sim\N(0,\sigma_j^2)$, $j=2,\dots,q$.

\begin{figure}
	\begin{center}
		\begin{tabular}{cc}
			$\quad \quad q=20$ & $\quad \quad q=40$ \\
			\includegraphics[scale=0.24]{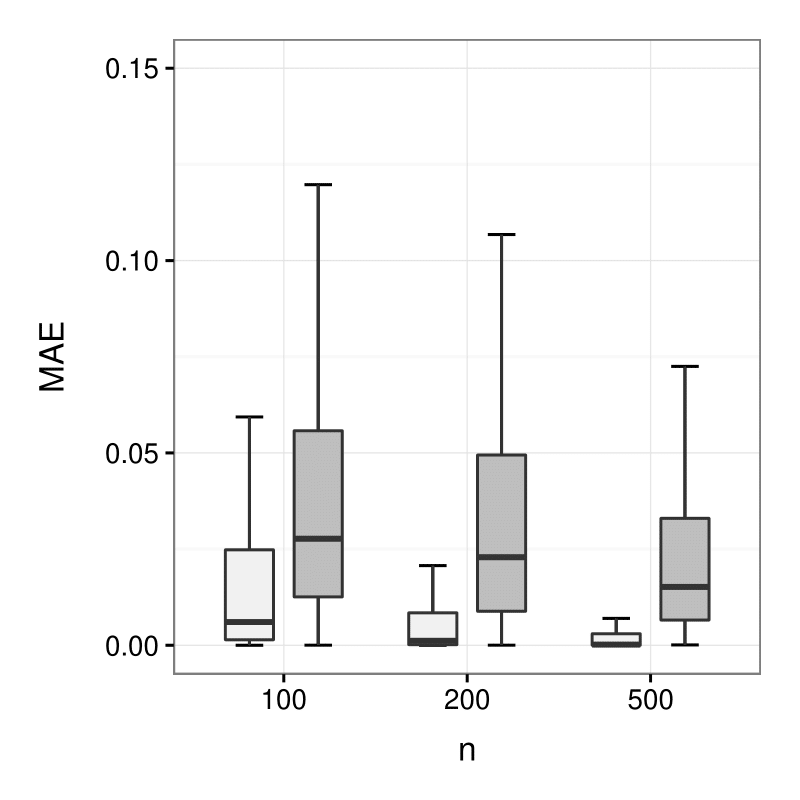} &
			\includegraphics[scale=0.24]{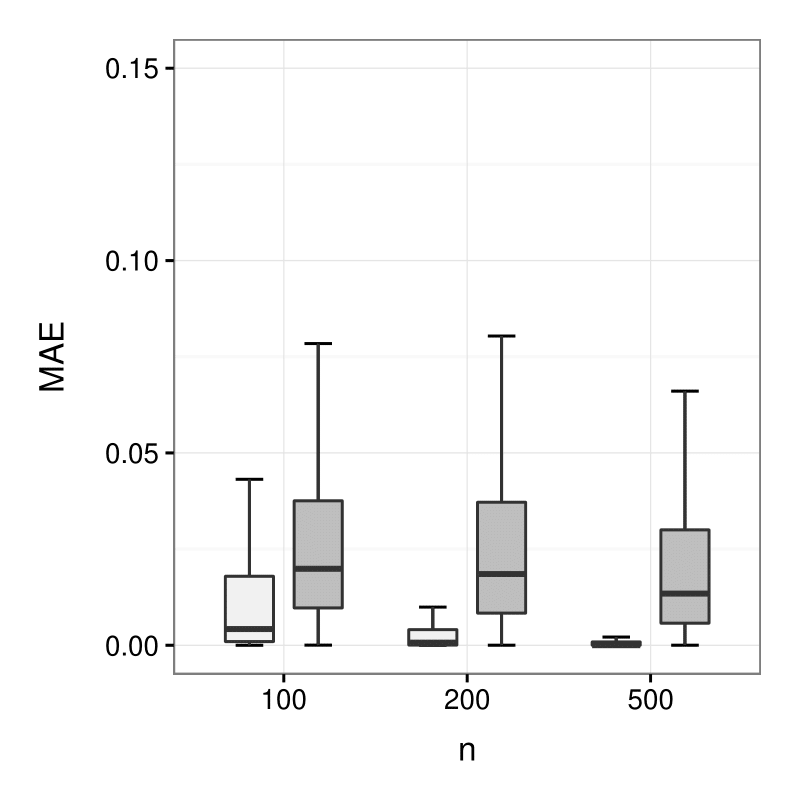}
		\end{tabular}
		\caption{\small Distribution over $40$ datasets and nodes $s \in \{2,\dots,q\}$ of the mean absolute error (MAE) of BMA estimates of true causal effects. Results are presented for two methods under comparison, our original DAG-probit method (light grey) and the \textit{naive} DAG-based approach (dark grey), for each combination of number of nodes $q\in\{20,40\}$ and sample size $n\in\{100,200,500\}$.}
		\label{fig:simulations:cfr}
	\end{center}
\end{figure}

We then construct simulation scenarios as detailed in Section 6 of our paper.
Results are summarized in the box-plots of Figure \ref{fig:simulations:cfr}, where we report the distribution of the Mean Absolute Error (MAE, constructed across the $40$ DAGs and nodes $s=2,\dots,q$) under the two strategies as a function of $n$.
As expected, our original method which fully accounts for the uncertainty on the DAG generating model, outperforms the alternative approach which is instead based on a fixed DAG not accounting for dependencies among variables.
Indeed, it appears that the causal effect estimate significantly deviates from the true causal effect even for large sample sizes.

\section{Computational time}

In this section we investigate the computational time of our method as a function of the number of variables $q$ and sample size $n$.
The following plots summarize the behavior of the running time (averaged over 40 replicates) \textit{per} iteration, as a function of $q \in \{5,10,20,50,100\}$ for $n = 500$, and as a function of $n\in\{50,100,200,500,1000\}$ for $q=50$.
Results were obtained on a PC Intel(R) Core(TM) i7-8550U 1,80 GHz.

\begin{figure}
	\begin{center}
		\begin{tabular}{c}
			\includegraphics[scale=0.27]{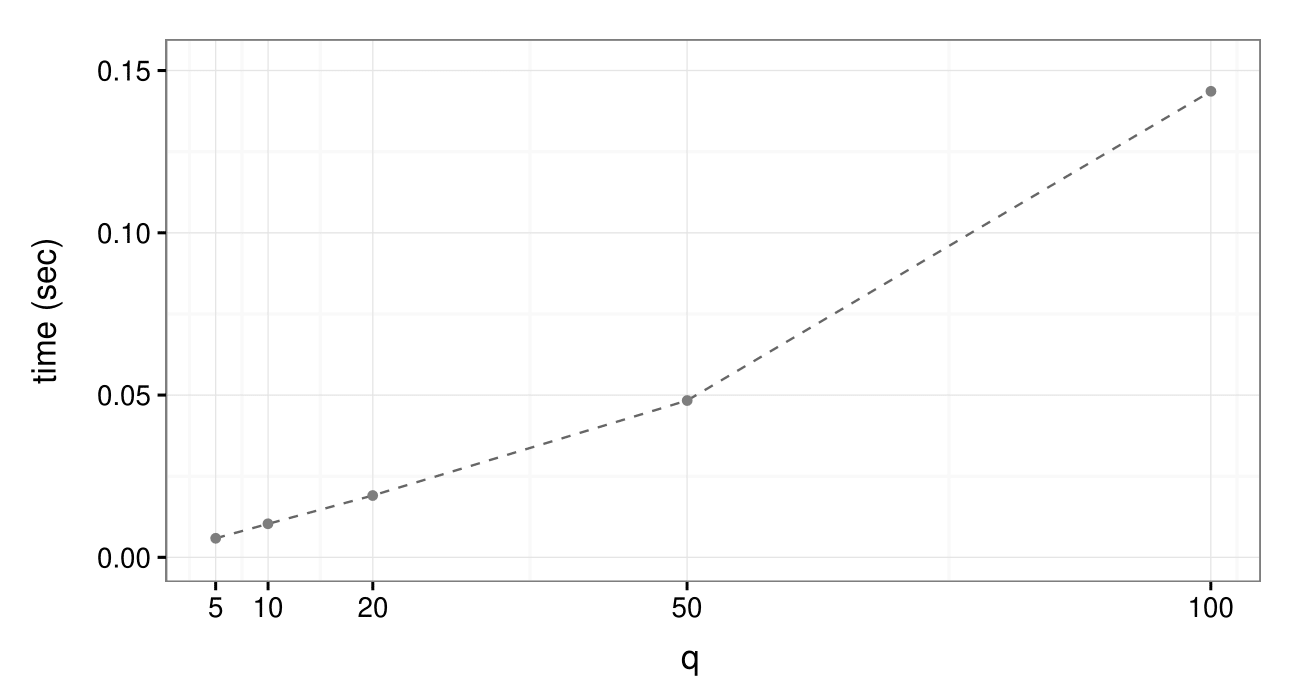} \\
			\includegraphics[scale=0.27]{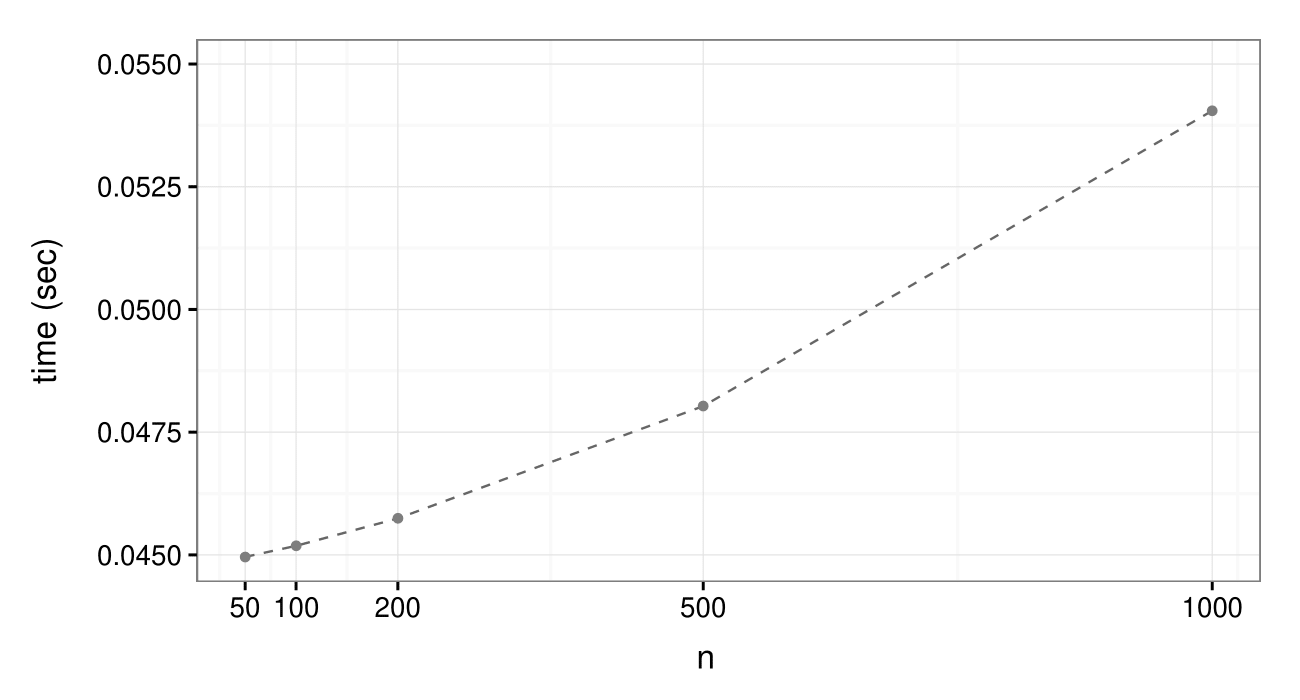}
		\end{tabular}
		\caption{\small Computational time (in seconds) \textit{per} iteration, as a function of the number of variables $q$ for fixed $n=500$ (upper plot) and as a function of the sample size $n$ for fixed $q=50$ (lower plot), averaged over 40 simulated datasets.}
		\label{fig:computational_time}
	\end{center}
\end{figure}

\bibliographystyle{biometrika}
\bibliography{biblio}